\documentclass[11pt,aps,prd,preprintnumbers,nofootinbib,showkeys,superscriptaddress]{revtex4-2}
\pdfoutput=1
\usepackage{float}
\usepackage{cancel}
\usepackage{amsfonts}
\usepackage{amsmath}
\usepackage{amssymb}
\usepackage{color}
\usepackage{tikz}
\usepackage{graphicx}
\usepackage{mathrsfs}
\usepackage{epsfig}
\usepackage{hyperref}
\usepackage{bm}% bold math
\usepackage{multirow} 
{}
{}

\def \nchi0{\widetilde\chi^0}

\newcommand{\newc}{\newcommand}
\newc{\ba}{\begin{array}}
\newc{\ea}{\end{array}}
\newc{\bea}{\begin{eqnarray}}
\newc{\eea}{\end{eqnarray}}
\newc{\beastar}{\begin{eqnarray*}}
\newc{\eeastar}{\end{eqnarray*}}
\newc{\beq}{\begin{equation}}
\newc{\eeq}{\end{equation}}
\newc{\bestar}{\begin{equation*}}
\newc{\eestar}{\end{equation*}}
\newc{\ben}{\begin{enumerate}}
\newc{\een}{\end{enumerate}}

\begin{document}

\title{Radiative corrections to the direct detection of the Higgsino-(and Wino-)like 
neutralino dark matter: Spin-dependent interactions }
%Radiative Corrections to Aid the Direct Detection of the Higgsino-like (and Wino-like) Neutralino 
%Dark Matter: Spin-Dependent Interactions}

\author{Subhadip Bisal}
\email{subhadip.b@iopb.res.in}
\affiliation{ Institute of Physics, Sachivalaya Marg, Bhubaneswar, 751 005, India}

\affiliation{Homi Bhabha National Institute, Training School Complex, Anushakti Nagar, Mumbai 400 094, India}

\author{Arindam Chatterjee}
\email{arindam.chatterjee@snu.edu.in}
\affiliation{Shiv Nadar Institution of Eminence deemed to be University, Gautam Buddha Nagar, Uttar Pradesh, 201314, India}

\author{Debottam Das}
\email{debottam@iopb.res.in}
\affiliation{ Institute of Physics, Sachivalaya Marg, Bhubaneswar, 751 005, India}
\affiliation{Homi Bhabha National Institute, Training School Complex, Anushakti Nagar, Mumbai 400 094, India}

\author{Syed Adil Pasha}
\email{sp855@snu.edu.in}
\affiliation{Shiv Nadar Institution of Eminence deemed to be University, Gautam Buddha Nagar, Uttar Pradesh, 201314, India}

\begin{abstract}
The lightest neutralino ($\tilde{\chi}_1^0$) is a promising dark matter (DM) 
candidate in the R-parity conserving minimal supersymmetric standard model 
(MSSM). In this work, we focus on dominantly Higgsino-like and Wino-like 
$\tilde{\chi}_1^0$ DM, with small admixtures of gauginos and Higgsinos, respectively. 
In particular, we explore large one-loop corrections to the $\tilde{\chi}_1^0
\tilde{\chi}_1^0Z$ vertex, which can significantly affect the estimation of the 
spin-dependent  $\tilde{\chi}_1^0$-nucleon scattering cross-section in the regions 
where such DM candidates are viable. We have used the on-shell renormalization 
scheme to estimate the relevant counterterm contributions. In the parameter 
region where $\tilde{\chi}_1^0$ is dominantly Higgsino-like, the radiative 
corrections (including the contributions from the respective counterterms) are 
substantial and can enhance the $\tilde{\chi}_1^0\tilde{\chi}_1^0Z$ vertex 
by up to $\sim 120\%$ for the benchmark scenarios we have considered. 
Further, for an almost pure Wino-like  $\tilde{\chi}_1^0$, the increment in the 
$\tilde{\chi}_1^0\tilde{\chi}_1^0Z$ vertex is up to $15\%$.  The corresponding 
cross-sections with the proton and the neutron can be changed by up to about
$50\%$. In addition, including the electroweak box diagrams, the cross-sections 
can be significantly enhanced, in particular, for the Wino-like $\tilde{\chi}_1^0$.
\end{abstract}

\maketitle
\section{Introduction}
\label{sec:intro}
There has been ample evidence of existence of non-luminous matter, i.e.,
dark matter (DM), in our Universe \cite{Bertone:2004pz,Bertone:2010zza,Bertone:2016nfn}. 
Cosmological observations suggest that DM constitutes about 26\% of the energy 
budget of the Universe \cite{Planck:2018vyg,Planck:2013oqw}. In the standard model 
of particle physics (SM) there is no suitable candidate for DM. However, several 
well-motivated extensions of SM incorporates suitable DM candidates 
\cite{Jungman:1995df,Bertone:2004pz}. 

Among the theoretical frameworks for beyond SM, supersymmetric extensions
have been well studied. In particular, the 
minimal supersymmetric standard model (MSSM) with conserved $R$-parity
has been widely studied in the literature. In this scenario, in the parameter regions 
where the lightest neutralino ($\tilde{\chi}_1^0$)  is the lightest odd particle 
under $R$-parity, the particle remains stable, and is a suitable candidate for DM. 
One of the prime motivations of the supersymmetric theories have been addressing
the naturalness concerns  \cite{Barbieri:1987fn, Ellis:1986yg, Chan:1997bi, 
Feng:2013pwa, Giudice:2013nak, Baer:2012cf, Mustafayev:2014lqa}. Stringent 
limits from the large hadron collider (LHC) on the masses of the heavy 
supersymmetric partners  \cite{ATLAS:2021yqv, ATLAS:2019wgx, ATLAS:2022zwa,
ATLAS:2022hbt,ATLAS:2021moa,CMS:2022sfi,CMS:2023xlp} have been in tension 
with naturalness requirements \cite{Baer:2012up, Baer:2012cf, Baer:2013ava, 
Mustafayev:2014lqa, Bae:2019dgg}. To quantify the naturalness, several measures 
of fine-tuning have been proposed 
in the literature, and such measures involve the Higgsino mass parameter $\mu$. 
One such measure, which uses the relevant parameters
at the electroweak scale, suggests that 
a rather small Higgsino mass parameter $\mu$ (of \(\mathscr{O}(100) \) 
GeV), is essential for electroweak naturalness.\footnote{Following the 
``electroweak" naturalness criteria, the fine-tuning measure has been estimated 
to be about $\mathcal{O}(10-100)$ assuming the masses of the third generation 
squarks and gluons in the ballpark of several TeV \cite{Baer:2012cf}.} In its minimal 
incarnation, assuming the gaugino mass parameters are very large 
$|\mu| \ll |M_1|, M_2$, a small $\mu$ parameter leads to Higgsino-like 
neutralino as the lightest stable particle (LSP). 

There have been numerous studies on the search of compressed neutralino 
spectra, including for Higgsino-like neutralinos at the LHC 
as well as in the direct and indirect DM detection experiments. The limits from 
the LHC on the compressed Higgsino spectrum, as mentioned above, is rather 
weak. Complimentary constraints on the $\mu$ parameter can be  obtained 
with Higgsino-like neutralinos as DM candidates \cite{Baer:2013vpa, 
Chakraborti:2017dpu,Dessert,Martin:2024ytt}, from the direct \cite{LZ:2022lsv, XENON:2023cxc, 
PandaX-II:2020oim, PICO-60} and indirect \cite{MAGIC:2016xys,
Fermi-LAT:2016uux,FermiLAT} searches for DM.\footnote{Even for a Bino-dominated LSP
accompanied with a Higgsino component, indirect searches may become important in 
particular context \cite{Chattopadhyay:2024qgs}.}

The prospects of direct detection of such neutralino DM critically depend on 
the composition of the $\tilde{\chi}_1^0$. For a pure Higgsino-like $\tilde{\chi}_1^0$, 
the tree-level interaction strength with the Higgs bosons, as well as with $Z$ boson, 
which dictate the spin-independent and spin-dependent $\tilde{\chi}_1^0$-nucleon 
cross-sections  repsectively, remain suppressed. Thus the radiative corrections 
to the respective vertices play an important role in determining the prospects of the 
direct detection. The importance of the radiative corrections {in the context 
of the spin-independent as well as the spin-dependent direct searches} have been 
studied in the literature in Refs.\cite{Drees:1996pk,Hisano:2004pv,Hisano:2011cs,
Hisano:2012wm,Bisal:2023fgb,Bisal:2024iar}.\footnote{Some important radiative 
corrections to the direct detection process and the relic abundance in the context 
of neutralino DM have been studied in Refs.\cite{Baro:2007em, Baro:2009na, 
Chatterjee:2012hkk, Harz:2014gaa,Harz:2014tma,Klasen:2016qyz, Harz:2023llw, 
Bisal:2023iip}.} In this article, we focus on some dominant radiative 
corrections which contribute to the spin-dependent direct detection cross-sections 
of the Higgsino-like DM. In particular, we evaluate the corrections to the 
$\tilde{\chi}_1^0 \tilde{\chi}_1^0 Z$ vertex, which affects the spin-dependent 
$\tilde{\chi}_1^0$-nucleon cross-section. 
Suitable versions of the on-shell renormalization scheme \cite{Fritzsche:2002bi, 
Baro2009,Chatterjee:2011wc} have been used to renormalize the chargino-neutralino 
sector  and contributions from the relevant vertex counterterms have been computed. 
In this context, we recall that, in Ref.\cite{Drees:1996pk} some dominant 
contributions to the $\tilde{\chi}_1^0 \tilde{\chi}_1^0 Z$ 
vertex corrections from the loops involving the 3rd generation quark-squarks have 
been considered and results were presented for rather light Higgsino-like 
states, approximately accounting for the effect of the gaugino-Higgsino mixing. The 
contributions from the loops involving the other (s)particles have not been considered. 
Additionally, in Refs.\cite{Hisano:2004pv,Hisano:2011cs,Hisano:2012wm} 
important radiative corrections to the 
$\tilde{\chi}_1^0 \tilde{\chi}_1^0 Z$ vertex involving the electroweak gauge 
bosons, as well as contributions from the electroweak box diagrams  
have been considered for Higgsino-like and Wino-like neutralino dark matter in the 
limit of respective pure gauge eigenstates.  However, the sfermions and fermions 
were assumed to be heavy and thus, their contributions have been ignored.
The present work takes into account the loop contributions from 
all the sfermions and fermions, as well as, gauge bosons, Higgs bosons and 
chargino-neutralinos respecting the relevant experimental constraints. Further, 
as mentioned, full vertex renormalization, including contributions 
from the vertex counterterms 
in the on-shell renormalization scheme, have been taken into account. The radiative 
corrections are then used to  estimate the spin-dependent cross-section. This is 
performed by implementing these corrections in the publicly available package 
\texttt{micrOMEGAs} \cite{Belanger:2010pz,Belanger:2013oya}. Further, 
significant contributions from the  electroweak box diagrams to the spin-dependent 
$\tilde{\chi}_1^0$-nucleon scattering have been incorporated following 
Refs.\cite{Hisano:2004pv,Hisano:2011cs,Hisano:2012wm}.  While the main focus 
of this work is on Higgsino-like $\tilde{\chi}_1^0$, a similar study on the 
Wino-like $\tilde{\chi}_1^0$ have also been pursued in this context of 
spin-dependent interactions. In case of Wino-like $\tilde{\chi}_1^0$, radiative 
corrections to the $\tilde{\chi}_1^0 \tilde{\chi}_1^0 Z$ vertex from loops involving 
gauge bosons, neutralinos and charginos, and  from the (s)fermions have 
been considered. As before, the vertex counterterms have been estimated 
using the on-shell renormalization scheme. Finally, dominant contributions 
from the electroweak box diagrams have been included following 
Refs.\cite{Hisano:2010fy,Hisano:2011cs}. As in the previous case, 
the radiative corrections are then  implemented in the publicly available 
package \texttt{micrOMEGAs} \cite{Belanger:2010pz,Belanger:2013oya} 
and their effect on the spin-dependent interactions have been discussed.\footnote{It
is worth noting that, the Wino mass parameter $M_2$ can be 
smaller than $|\mu|, |M_1|$  in specific realizations of the 
high scale models, in particular, where anomaly mediated supersymmetry breaking (AMSB) may 
be realized \cite{Randall:1998uk,Giudice:1998xp}. Thus, in such scenarios, 
$\tilde{\chi}_1^0$ can be Wino-like.}

The thermal relic density of $\Omega_{\rm DM} h^2 \approx 0.12$  in the early 
universe is achieved for a Higgsino-like $\tilde{\chi}_1^0$ with the Higgsino mass 
parameter $\mu$ around 1 TeV, and for a Wino-like $\tilde{\chi}_1^0$  mass 
parameter $M_2$ of about 2 TeV\footnote{Considering Sommerfeld enhancement 
leads to an enhanced (co-)annihilation cross-section for rather heavy 
($\mathcal{O}(1 $TeV)) neutralino DM \cite{Hisano:2006nn, Beneke:2014hja,Hisano:2004ds, 
Mohanty:2010es,Beneke:2019qaa}. For the Wino-like $\tilde{\chi}^0_1$ the correct relic 
density is achieved at $M_2 \sim 2.7-3$ TeV. 
%We have not considered this effect in the analysis of the 
%Higgsino-like and Wino-like $\tilde{\chi}^0_1$.
}, the details also depend on the other 
supersymmetric sparticle spectra.\footnote{These mass scales can be brought down if we 
consider coannihilation effects, see, e.g. \cite{Chakraborti:2015mra}. 
Moreover, there are scenarios discussed in literature for non-thermal production of 
DM \cite{Aparicio:2016qqb}. Further, there may exist other components of dark 
matter, i.e., axions, etc. which may resolve the under-abundance of lower mass 
Higgsino or Wino-like $\tilde{\chi}_1^0$ DM \cite{Bae:2013hma}.} Note that in case 
of $\tilde{\chi}_1^0$ constituting only a fraction of the DM content of the universe, 
the direct detection bounds are relaxed in the same proportion. In this article, we 
will not concern ourselves with satisfying the thermal relic abundance of the early 
universe.  
 
This article is organized as follows. 
In section \ref{sec:outline}, we present the theoretical outline of the chargino-neutralino 
sector and the tree-level coupling of the neutralino with the $Z$ boson. In section 
\ref{sec:DD}, we motivate the Higgsino-like and Wino-like LSP and their direct 
detection via underground and ground based detectors. Section \ref{sec:results} 
contains the results of our work for both Higgsino-like as well as Wino-like case and 
their discussion. In section \ref{sec:Conclusion}, we conclude our work.
 
 %%%%%%%%%%%%%%%%%%%%%%%%%%%%%%%%%%%%%%%%%%%%%%%%%%%%
\section{The Framework}
\label{sec:outline}

In this section, we begin by discussing the chargino-neutralino sector in the MSSM; 
see, e.g., \cite{Drees:2004jm}, for a review. In particular, we focus on the parameter region 
with a Higgsino-like and a Wino-like lightest neutralino. Subsequently, we discuss the 
generalities of spin-dependent $\tilde{\chi}_1^0$-nucleus scattering, and emphasize 
the implications of radiative corrections to the $\tilde{\chi}_1^0 \tilde{\chi}_1^0 Z$ 
vertex on the respective cross-section. 

\subsection{The Chargino-Neutralino Spectrum}

The tree-level mass Lagrangian for the charginos can be expressed in the gauge eigenbasis in terms of the Weyl spinors (the charged Winos \(\widetilde{W}^+\) and \(\widetilde{W}^-\), and the Higgsinos \(\tilde{h}^\pm_i\) for \(i \in \{1, 2\}\)) with \(\psi^+ = (\widetilde{W}^+, \tilde{h}^+_2)\) and \(\psi^- = (\widetilde{W}^-, \tilde{h}^-_1)\), as \cite{Drees:2004jm}:
\begin{align}
-\mathcal{L}^c_{\mathrm{Mass}} = \psi^{- T}M^c\psi^+ + H.C.
\end{align}
here $M^c$ is the mass matrix written as: 
\begin{align}
M^{c}= \left( \begin{array}{cc}
M_{2} & \sqrt{2} M_W\sin \beta \\
\sqrt{2} M_W \cos \beta & \mu \\
\end{array} \right).
\end{align}

In the above expression, \(M_2\) and \(\mu\) represent the supersymmetry-breaking $SU(2)$ Wino mass 
parameter and the supersymmetric Higgsino mass parameter, respectively; \(M_W\) 
denotes the mass of the \(W\) boson, and \(\tan \beta\) is the ratio of the vacuum 
expectation values (\textit{vevs}) of the up-type and down-type CP-even neutral Higgs 
bosons. The chargino mass matrix \(M^c\) can be diagonalized through a bi-unitary transformation 
with the help of unitary matrices \(U\) and \(V\) to get:
\beq \label{mcd}
M^{c}_{D} = U^*M^{c}V^{-1} = {\rm Diagonal} (m_{\tilde{\chi}_{1}^{+}}
~m_{\tilde{\chi}_{2}^{+}}) , 
\eeq

The mass eigenstates are ordered as: \( m_{\tilde{\chi}_1^+} \leq m_{\tilde{\chi}_2^+} \). For the 
charginos \((\tilde{\chi}_i^+ \text{ with } i \in \{1, 2\})\), the left-handed and right-handed 
components of these mass eigenstates can be expressed as:
\beq \label{ec}
\bold{P_L} \tilde \chi^{+}_i = V_{ij} \psi^+_j,~~ \bold{P_R} \tilde \chi^+_i = U^*_{ij}
  \overline{\psi^-_j}\,,
\eeq
where $\bold{P_L}$ and $\bold{P_R}$ are the standard projection operators, $\overline{\psi^-_j} = \psi^{- 
\dagger}_j$, and $j$ is summed over.

The neutralino states consists of the Bino (\(\widetilde{B}^0\)), the neutral Wino (\(\widetilde{W}^3\)), and the down-type and up-type neutral Higgsinos (\(\tilde{h}^0_1\) and \(\tilde{h}^0_2\), respectively). Written in the gauge eigenbasis, \(\psi^0 = (\widetilde{B}, \widetilde{W}^3, \tilde{h}^0_1, \tilde{h}^0_2)\), the Lagrangian for the mass term assumes
the form \cite{Drees:2004jm}:
\beq
-\mathcal{L}^{n}_{\rm Mass}  =  \frac{1}{2} \psi^{0 T} M^n
\psi^{0} + {\rm h.c.}.  
\eeq
$M^{\rm n}$ is the neutralino mass matrix given by,
\beq \label{mn}
M^{n} =  \left( \begin{array}{cccc}
M_{1} & 0 & -M_{Z}s_W c_\beta & M_{Z}s_W s_\beta\\
0 &  M_{2} & M_{Z}c_W c_\beta & -M_{Z}c_W s_\beta \\
-M_{Z}s_W c_\beta & M_{Z}c_W c_\beta & 0 & -\mu \\ 
M_{Z}s_W s_\beta & -M_{Z}c_W s_\beta & -\mu & 0 \end{array} \right)\,.
\eeq

In Eqs.(\ref{mcd}) and (\ref{mn}), $ s_W $, $s_\beta $, $c_W$, and $c_\beta$ represent 
$\sin \theta_W$, $\sin \beta$, $\cos \theta_W$, and $\cos \beta$ respectively, where 
$\theta_W$ is the weak mixing angle. $M_Z$ denotes the mass of the $Z$ boson, and 
$M_1$ is the supersymmetry breaking $ U(1)_Y$ gaugino (Bino) mass parameter. In order 
to determine the masses of the neutralinos, the matrix $M^n$ can be diagonalized by a 
unitary matrix $N$ as follows:
\beq \label{mnd}
M^{n}_{D}=N^* M^{n} N^{-1}= {\rm  Diagonal} (m_{\tilde{\chi}^{0}_{1}} ~
m_{\tilde{\chi}^{0}_{2}} ~m_{\tilde{\chi}^{0}_{3}} ~m_{\tilde{\chi}^{0}_{4}})
\eeq

The mass eigenstates (\(\tilde{\chi}^0_i\)) are arranged in increasing order of their corresponding
mass eigenvalues: \(m_{\tilde{\chi}^0_1} \leq m_{\tilde{\chi}^0_2} \leq m_{\tilde{\chi}^0_3} \leq 
m_{\tilde{\chi}^0_4}\) (in absolute value). These eigenstates are charge conjugates of themselves, 
i.e., \(\tilde{\chi}^{0C}_i = \tilde{\chi}^0_i\), where the superscript \(C\) denotes charge 
conjugation. For the Majorana neutralinos $\tilde{\chi}^0_i  (i \in \{1, 2, 3, 4\})$, the 
left-handed components of these mass eigenstates can be expressed as:
\beq
\bold{P_L} \tilde \chi^{0}_i = N_{ij}  \psi_j^{0},
\eeq
where $j$ is summed over again and $N_{ij}$ refers to the $\{i,j\}^{th}$ element of the 
matrix $N$ in Eq.(\ref{mnd}).

Analytical expressions for the mass eigenvalues of charginos and neutralinos are available in 
the literature \cite{Barger1994,Kheishen,Choi:2001ww, Bertone:2004pz}. But, we can numerically 
estimate these eigenvalues (esp. for neutralinos) in a straightforward and practical manner.

In the present context, we state the approximate mass eigenvalues  for the Higgsino-like 
and the Wino-like neutralinos, in the limit of negligible mixing. In the parameter 
region where the Higgsino mass parameter is notably smaller compared to the gaugino  mass 
parameters, i.e., \( |\mu| \ll |M_1|, M_2 \), the masses of the light Higgsino-like particles can 
be approximated as \cite{Drees:1996pk,Giudice:1995np}:
\bea
m_{\tilde\chi^{\pm}_1} & = & |\mu |\left(1- \frac{M_W^2 \sin2\beta}{\mu  M_2}\right) + 
\mathcal{O}(M_2^{-2})+ {\rm rad. corr.} \nonumber\\
m_{\tilde\chi^{0}_{a,s}} & = & \pm \mu - \frac{M_Z^2}{2}(1\pm \sin2\beta)
\left(\frac{s_W^2}{M_1}+\frac{c_W^2}{M_2}\right) + {\rm rad. corr.} 
\label{eq:mHiggsino}
\eea

In the expression above, subscripts $a(s)$ denote anti-symmetric (symmetric) combinations 
of up-type ($\tilde{h}^0_2$) and down-type ($\tilde{h}^0_1$) Higgsinos constituting the 
respective mass eigenstates. The symmetric and anti-symmetric states refer to Higgsino-like 
states composed without and with a relative sign between $N_{i3}$ and $N_{i4}$, 
respectively.\footnote{It has been noted in the literature that, due to mixing effects, 
the mass differences $\Delta 
m_1 = m_{\tilde{\chi}_1^{\pm}} - m_{\tilde{\chi}^0_1}$ may become very small in certain 
regions of the parameter space \cite{Kribs:2008hq, Han:2014kaa, Barducci:2015ffa, 
Chatterjee:2017nyx}. In this context, we focus on mass differences \(\Delta m_1, \Delta 
m_2 \gg \mathcal{O}(1 \text{ MeV})\), where $\Delta m_2 = m_{\tilde{\chi}_2^{0}} - 
m_{\tilde{\chi}^0_1}$. Therefore, as discussed in the next section, only 
the elastic scattering of $\tilde{\chi}^0_1$ with nucleons will be relevant in direct detection 
experiments.}

In the parameter region where the Wino mass parameter is much smaller compared to the 
Bino and Higgsino mass parameters, i.e., $ M_2 \ll |M_1|, |\mu|$ , the masses of the  Wino-like 
states can be approximated as follows \cite{Hisano:2004pv}, 
\bea
m_{\tilde\chi^{\pm}_1} & = & M_2 + \frac{M_W^2 }{M_2^2- \mu^2}\left(M_2+ \mu \sin2\beta \right) + {\rm rad. corr.} +... \nonumber\\
m_{\tilde\chi^{0}_{1}} & = & M_2 + \frac{M_W^2 }{M_2^2-\mu^2}\left(M_2+ \mu \sin2\beta \right) + {\rm rad. corr.}+...
\label{eq:mWino}
\eea
Although at the tree-level the mass difference $\Delta m_1$ is very small, it has been 
noted that sizable mass splitting of $\mathcal{O}(100)$ MeV can be achieved
via the radiative corrections in the limit of negligible Higgsino-gaugino mixing. 
Thus, as stated in the previous case,  in the next section we only focus on the
the elastic scattering of $\tilde{\chi}^0_1$ with nucleons which will be relevant in 
direct detection experiments.

\subsection{Dark Matter direct detection : spin-dependent interactions between $\tilde{\chi}_1^0$ and nucleus}
 \label{sec:DD}
\subsubsection{Generalities of spin-dependent interactions}
We briefly review the generalities of the spin-dependent DM-nucleon interactions 
in the context of $\tilde{\chi}_1^0$ DM following Refs.\cite{Goodman:1984dc,Griest1, Drees1,Jungman:1995df}. The effective Lagrangian for the $\tilde{\chi}_1^0$-nucleon 
spin-dependent interaction is given by \cite{Goodman:1984dc,Griest1,Griest:1988ma,Jungman:1995df}, 
\begin{align}\label{eq:LeffN}
\mathcal{L}_{\mathrm{eff}} \supset  g_N \bar{\chi} \gamma_{\mu} \gamma_5 \chi \bar{\psi}_N 
\gamma^{\mu} \gamma_5 \psi_N,
\end{align}
where $g_N$ is the effective coupling, $N \in \{n,p\}$ denotes a neutron or a proton and 
$\psi_N$ denotes the respective quantum field. At the parton level, the relevant 
effective Lagrangian describing the $\tilde{\chi}^0_1$-quark interactions is given by 
\cite{Goodman:1984dc,Griest1, Griest:1988ma,Drees1, Jungman:1995df}, 
\begin{align}\label{eq:Leffq}
\mathcal{L}^{q}_{\rm eff, SD} \supset d_q \overline{\tilde{\chi}^0_1}\gamma^{\mu}\gamma_5 
\tilde{\chi}^0_1 \overline{\psi}_q \gamma_{\mu}\gamma_5 \psi_q ,
\end{align}
where $q$ denotes the (light) quark under consideration, $\psi_q$ denotes the 
respective quantum field and $d_q$ denotes the respective effective coupling. 
$g_N$ can be expressed in terms of the effective interaction strength $d_q$ as 
follows \cite{Goodman:1984dc,Griest1,Jungman:1995df}, 
\beq
g_N = \sum_{q= u,d,s} d_q \Delta q_{N}.
\eeq
In the above expression $\Delta q_{N}$, which is a measure of the fraction of spin 
of nucleon $N$ contributed by the constituent quark $q$, is related to  the matrix 
element of the respective quark axial-vector current, which is given by, 
\beq
\langle N | \bar{\psi}_q \gamma^{\mu} \gamma_5 \psi_q | N \rangle =  
2 s^{\mu}_N \Delta q_{N}.
\eeq
In the above equation, $d_q$ denotes the parton level effective coupling involving 
axial vector currents of quarks and $\tilde{\chi}_1^0$. Further, $s^{\mu}_N$ denotes 
the spin of the nucleon $N$. $\Delta q^{N}$ are extracted from experimental data, 
see, e.g., Ref.\cite{HERMES:2006jyl}.\footnote{We follow the publicly available package 
\texttt{micrOMEGAs} in this regard, which adopted these from Ref.\cite{HERMES:2006jyl}. 
Further details in this regard can be found in Appendix A.}

Note that, for the nucleon at rest, the $\mu =0$ component of the axial-vector current
vanishes. To obtain the spin-dependent $\tilde{\chi}_1^0$-nucleus scattering 
cross-section, next, the matrix element of the nucleon spin operators in the nuclear 
state is estimated. At zero momentum transfer, it is given by the expectation value 
of the spin content of the protons and neutrons in the nucleus $\tilde{N}$, 
denoted by $\langle S_p\rangle = \langle \tilde{N}| S_p|\tilde{N} \rangle$ and  
$\langle S_n\rangle = \langle \tilde{N}| S_n|\tilde{N} \rangle$ respectively  \cite{Engel:1992bf,Jungman:1995df}. Further, a suitable form factor $S(k)$ (where $k = |\vec{k}|$ 
gives the magnitude of the momentum transfer) is introduced to account for non-zero 
momentum transfer. In terms of the isoscalar and the isovector parameters 
$a_0 = (\sqrt{2} G_F)^{-1} (g_p + g_n)$ and $a_1 = (\sqrt{2} G_F)^{-1} (g_p-g_n) $, 
where $G_F = \frac{g_2^2}{8 m_W^2}$, $m_W$ denotes the mass of $W$ boson and  
the subscripts $p,~n$ denote proton and neutron respectively. The form factor $S(k)$, 
then can be expressed as,
\beq
S(k)= a_0^2 S_{00}(k) + a_1^2 S_{11}(k)+a_0a_1 S_{01}(k),
\eeq
where, the interference of the isoscalar and the isovector terms give rise to three 
independent form factors $S_{ij},~ i,j \in\{0,1\}$ \cite{Engel:1992bf,Jungman:1995df}.\footnote{We 
follow the form factors implemented in the publicly available code 
$\texttt{micrOMEGAs}$ \cite{Belanger:2008sj}. For the present 
work, see also Refs.\cite{Engel:1992bf,Jungman:1995df,Bednyakov:2004xq}.}
Finally, in the limit of zero momentum transfer, the spin-dependent $\tilde{\chi}_1^0$- 
nucleus differential scattering cross-section ($\sigma^{\tilde{N}}_{0~\rm SD}$) can be 
expressed as \cite{Engel:1992bf,Ressell:1993qm, Jungman:1995df}
\beq
    \frac{d\sigma^{\tilde{N}}_{0~\rm SD}}{dk^2}= \frac{4}{\pi v^2} \big(g_p \langle S_p \rangle + g_n \langle S_n\rangle\big)^2  \frac{\big(J+1\big)}{J}.
\eeq
In the above expression, $v$ denotes the relative speed of the DM with respect to the 
target nucleus and $J$ denotes the total angular momentum of the nucleus.
The effect of non-zero momentum transfer, as usual, is captured by introducing a suitable 
form factor $\mathcal{F}^2(k) = \dfrac{S(k)}{S(0)}$, as discussed above. The respective 
differential cross-section is given by \cite{Jungman:1995df}, 
\beq
    \frac{d\sigma^{\tilde{N}}_{\rm SD}}{dk^2} = \frac{d\sigma^{\tilde{N}}_{0~\rm SD}}{dk^2} \frac{S(k)}{S(0)}.
\eeq
The total spin-dependent scattering cross-section, in the limit of zero momentum 
transfer is given by \cite{Jungman:1995df}, 
\beq
\sigma^{\tilde{N}}_{0~ \rm SD} = 4 m_{r}^2 v^2  \frac{d\sigma^{\tilde{N}}_{0~ \rm SD}}{dk^2} = \frac{16}{\pi}m_r^2  \frac{(J+1)}{J} \big(g_p \langle S_p \rangle + g_n \langle S_n\rangle\big)^2 
\eeq
where, $m_r = \dfrac{m_{\tilde{\chi}_1^0}m_T}{m_{\tilde{\chi}_1^0}+m_T}$
denotes the reduced mass of the target nucleus $(\tilde{N})$ (with mass $m_T$) 
and  $\tilde{\chi}_1^0$. In the following subsection, we discuss the relevant 
$\tilde{\chi}_1^0$-quark effective interaction within the framework of MSSM.

\subsubsection{Spin-dependent interaction at tree-level and at one loop : the case for Higgsino-like and Wino-like $\tilde{\chi}_1^0$ DM}
As stated in the Introduction, in the present context we focus on the spin-dependent 
interactions between $\tilde{\chi}_1^0$ and nucleus. In particular, Higgsino-like and 
Wino-like $\tilde{\chi}_1^0$ states are of interest. The relevant effective 
Lagrangian describing the spin-dependent $\tilde{\chi}_1^0$-nucleon inetraction 
has been described already in Eq.(\ref{eq:LeffN}).  

At the level of partons, the tree-level contributions to the $\tilde{\chi}^0_1$-quark 
scattering comes from the $Z$-boson exchange ($t$-channel) process, as well 
as from the respective squark exchange ($s$-channel) processes \cite{Goodman:1984dc,Griest1,Griest:1988ma,Drees1,Jungman:1995df}. 
At the tree-level, the effective coupling $d_q$, as described in Eq.(\ref{eq:Leffq}), 
receieves contributions from the processes mentioned above. These contributions 
can be described as \cite{Drees1}
\begin{align}\label{eq:dqtree}
d_q^{\rm tree} = \dfrac{1}{4} \sum_{i=1}^{2} \frac{a_{\tilde{q}_i}^2+ b_{\tilde{q}_i}^2}
{m_{\tilde{q}_i}^2-(m_q+m_{\tilde{\chi}_1^0})^2} -\frac{g_2^2}{4 c_W^2 M_Z^2}\mathcal{N}_{11}^RT_{3q}
\end{align}
where $q$ denotes the light quarks (u,d,s) with mass $m_q$, $T_{3q}$ is the respective
weak isospin, $g_2$ denotes the $SU(2)_L$ gauge coupling and 
$\mathcal{N}_{11}^R$ denotes the coupling of neutralinos with $Z$ boson, 
which will be described below. The first term involves a sum over the two 
squark mass eigenstates $\tilde{q}_i,~i\in \{1,2\}$ of the same flavor as $q$, 
where $m_{\tilde{q}_i}$ denotes their respective masses. The expressions 
for $a_{\tilde{q}_i},~ b_{\tilde{q}_i}$ can be found in Ref.\cite{Drees1}.

For the $Z$ boson mediated process,  the relevant vertex involving $\tilde{\chi}^0_1$ is given by \cite{Haber:1984rc,Drees:2004jm},
\begin{align}\label{eq:LchiZ}
\mathcal{L}_{\tilde{\chi}^0_1\tilde{\chi}^0_1 Z} = \left(\dfrac{g_2}{2 c_W}\right) \overline{\tilde{\chi}^0_1} \gamma^{\mu} (\mathcal{N}_{11}^L \bold{P_L} + \mathcal{N}^R_{11} \bold{P_R})\tilde{\chi}^0_1 Z_{\mu},
\end{align}
where,
\begin{subequations}
\begin{align}
\mathcal{N}_{11}^L &= -\dfrac{1}{2} (|N_{13}|^2 - |N_{14}|^2), \\
\mathcal{N}_{11}^R &= -(\mathcal{N}_{11}^L)^*.
\end{align}
\end{subequations}

In the expressions above $N_{13} $ and $ N_{14}$ denote the down-type and 
up-type Higgsino compositions, respectively, in the mass eigenstate 
$\tilde{\chi}^0_1$. 

For Higgsino-like $\tilde{\chi}^0_1$, we have $|\mu| \ll |M_1|, M_2$.
Thus, the gaugino fraction is very small compared to the Higgsino fraction. Further, 
the mass eigenstate is composed of (approximately) equal proportions of 
down-type and up-type neutral Higgsino; thus $|N_{13}|^2- 
|N_{14}^2|$ is generally small, suppressing the $Z$ mediated contribution. Further, 
as the squark mediated contributions are proportional to the respective Yukawa 
couplings, the respective contributions are sub-dominant. Also, as we will 
discuss in Sec.\ref{sec:results}, in our work we have 
considered the (first two generation of) squarks to be very heavy. Therefore, 
the respective contributions remain sub-dominant. For a Wino-like $\tilde{\chi}^0_1$, 
$M_2$ is much smaller than  $|\mu|$, $|M_1|$. Consequently, the Bino and the 
Higgsino fractions are very small as compared to the Wino fraction. As $N_{13}$ 
and $N_{14}$ are very 
small, the tree-level vertex involving $Z$ boson is suppressed in this scenario.\footnote{ 
The approximate analytical expressions of the tree-level 
$\tilde{\chi}^0_1 \tilde{\chi}^0_1 Z$ can be found in Ref.\cite{Hisano:2004pv}. 
In this work numerical values for the mixing matrices and the tree-level 
vertices have been used to precisely estimate the respective contributions 
for various benchmark scenarios. }
The $SU(2)_L$ doublet squark mediated processes, as described above, 
contribute to the spin-dependent cross-section. As we have assumed these 
squarks to be heavy in the present work, the contributions remain rather small.
It is worth noting that, due to small tree-level cross-section, the radiative 
corrections to these vertices  can be important and can play a crucial role in 
spin-dependent direct detection processes.

As already mentioned, because of the smallness of the 
tree-level contributions to the spin-dependent $\tilde{\chi}^0_1$-nucleus 
scattering process, it is important to consider radiative corrections. 
In the next section, we will discuss 
the radiative corrections to the $\tilde{\chi}^0_1 \tilde{\chi}^0_1 Z$ vertex, which 
contributes to the coefficient $d_q^{\rm tree}$, as mentioned above. Further, 
certain important contributions from the box diagrams involving the electroweak 
gauge bosons, which were considered in Refs.\cite{Hisano:2004pv,Hisano:2012wm}, 
will also be discussed in the subsequent subsection, as we have incorporated 
these corrections also for the numerical estimation of the spin-dependent 
 $\tilde{\chi}^0_1$-nucleus cross-sections.

\subsubsection{Radiative corrections : $\tilde{\chi}^0_1 \tilde{\chi}^0_1 Z$ vertex}

In this section we focus on dominant radiative corrections to the 
$\tilde{\chi}^0_1 \tilde{\chi}^0_1 Z$ vertex and the relevant counterterms. As discussed 
in the previous section, the $Z$-boson mediated $\tilde{\chi}^0_1$-quark scattering 
processes contribute substantially to the spin-dependent direct detection. As the 
tree-level vertex for $\tilde{\chi}^0_1 \tilde{\chi}^0_1 Z$ is small for almost pure 
Higgsino-like and Wino-like neutralinos, the radiative corrections can be important 
for accurate estimation of the spin-dependent direct detection cross-section of 
$\tilde{\chi}^0_1$.

\begin{figure}
	\centering
	\includegraphics[scale=0.48]{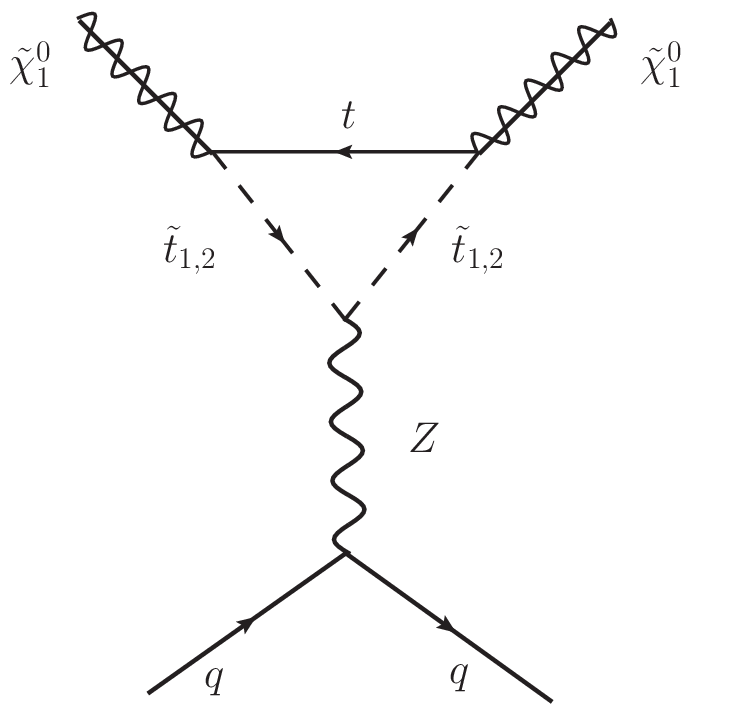}
	\includegraphics[scale=0.48]{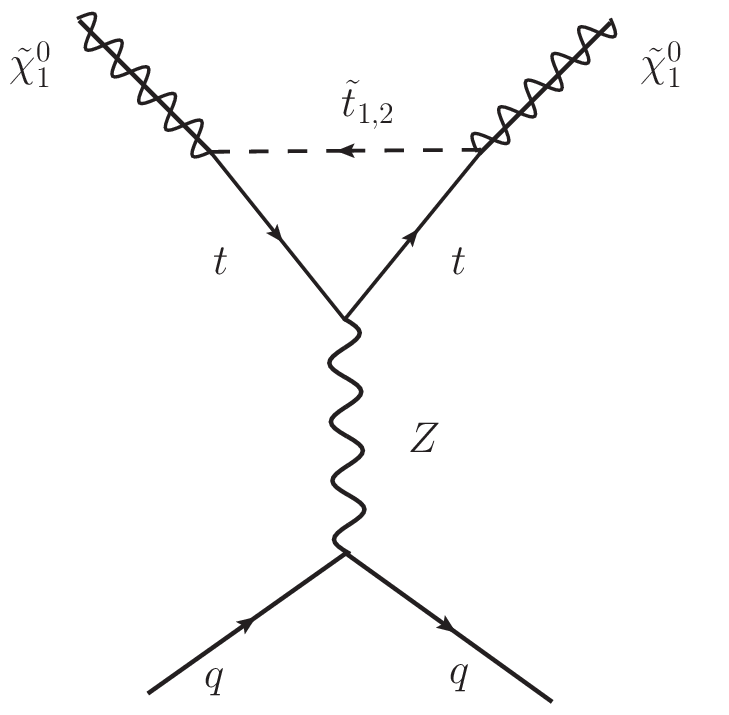}
	\caption{Feynman diagrams depicting  the contributions from the  third generation (s)quark 
	sector to the $\tilde{\chi}_1^0\tilde{\chi}_1^0Z$ vertex corrections. }
	\label{fig:vertexcorr1}
\end{figure}

The radiative corrections to the $\tilde{\chi}^0_1 \tilde{\chi}^0_1 Z$ vertex generally 
receive contributions from the gauge bosons, Higgs bosons, neutralinos and 
charginos, and also from fermion and sfermions running in the loops. {For 
the Higgsino-like $\tilde{\chi}_1^0$ scenario, due to the large Yukawa coupling,
the diagrams involving third-generation top quarks and stop squarks provide the most 
significant radiative contributions}. The respective diagrams have
been depicted in Fig.\ref{fig:vertexcorr1}.\footnote{The contributions from the third 
generation (s)quark loops only, in the context of an almost pure Higgsino-like neutralino, 
have been discussed in Ref.\cite{Drees:1996pk} using $\overline{\rm DR}$ renormalization 
scheme, and the effects in the direct detection event rate for very low $\tan\beta 
\simeq 1.5$ and rather light neutralino mass ($\simeq 70$ GeV), which are dominated by 
spin-independent cross-section, were presented. Our focus will be primarily on the 
implications of the $\tilde{\chi}^0_1 \tilde{\chi}^0_1 Z$ vertex corrections, as 
evaluated in the on-shell scheme, on the spin-dependent cross-section for both 
Higgsino-like and Wino-like neutralinos in the parameter region which are consistent 
with the current experimental constraints.} Further, there are sub-dominant 
contributions from the loops involving two neutralinos (charginos) and $Z$ ($W^{\pm}$) 
bosons. There is some cancellation between the loops involving $\tilde{\chi}_1^{+} W^{+} 
W^{+} $ and $\tilde{\chi}_1^{-} W^{-} W^{-} $ {(shown in 
Fig.\ref{fig:vertexcorrWW}(a))}. Thus, their contribution to the vertex correction 
remains insignificant, which was also pointed out in {Ref.\cite{Hisano:2011cs}.
This cancellation is exact when we take the pure Higgsino (and pure Wino) 
limit, but when there is a small admixture of Bino and Wino (Bino and Higgsino) in the
$\tilde{\chi}^0_1$, i.e., the case considered in this article, this is not an exact 
cancellation.}
Further, there is a partial cancellation between the loops involving $\tilde{\chi}_1^{+}
\tilde{\chi}_1^{+}  W^{+}$ and $\tilde{\chi}_1^{-}  \tilde{\chi}_1^{-}  W^{-} $ 
{(Fig.\ref{fig:vertexcorrWW}(b))}, thus 
effectively reducing their contributions. {Similar to the previous case, this 
cancellation is exact in the pure Higgsino (and pure Wino) limit. This cancellation is also
discussed in Ref.\cite{Hisano:2011cs}.}
Note that, in the limit of very heavy gaugino mass parameters, i.e. $|M_1|, M_2 \gg 
|\mu|$, the dominant contributions from the triangle loops involving two 
neutralinos and a $Z$ boson are also suppressed by at least one factor of 
$|N_{i3}|^2-|N_{i4}|^2$, $i \in \{1,2\}$. The analytical expressions for different loop 
diagrams may be found in Appendix \ref{sec:App-B}. However, as we will discuss in 
Sec.\ref{sec:results}, the contribution, for $|M_1|, M_2 \lesssim 5$ TeV, and for $|\mu|
\simeq 1$ TeV, their contribution can be about a few percent compared to the 
dominant loops. 

For a Wino-like neutralino ($M_2 \ll |M_1|, |\mu|$), on the other hand, contributions 
to the $\tilde{\chi}^0_1 \tilde{\chi}^0_1 Z$ vertex from the loops involving left-(s)fermions 
are sizable. However, there are cancellations between the loops involving two 
fermions and one fermion and the loops involving two sfermions and one fermion of the 
same flavor. Further, contributions from the loop with two $W^{+}$ bosons and one 
(Wino-like) chargino, and from the loop with the respective anti-particles largely cancel, 
see also Ref.\cite{Hisano:2004pv}. Thus, the total contribution from the loops, as well 
as from the counterterms is not as large as in the Higgsino case.  

It is necessary to use a renormalization procedure for the estimation of radiative 
corrections. We use an appropriate variant of the on-shell renormalization scheme 
for the chargino-neutralino sector \cite{Fritzsche:2002bi,Baro2009,Chatterjee:2011wc} 
to renormalize the chargino-neutralino sector. The respective 
$\tilde{\chi}^0_1 \tilde{\chi}^0_1 Z$ vertex counterterms have been considered, 
following the implementation in Ref.\cite{Fritzsche:2013fta}. As we will discuss 
in the section \ref{sec:results}, there are generally large cancellations between the 
loop corrections and the counterterm contributions to the 
$\tilde{\chi}^0_1\tilde{\chi}^0_1 Z$ vertex. To obtain the vertex counterterm, one 
expresses the ``bare" and the renormalized neutralino as

\beq \label{eq:RenChi}
\tilde{\chi}_i^{0~\rm bare} = \big(\delta_{ij} + \dfrac{1}{2} \delta {\rm Z}^L_{ij} \bold{P_L} + \dfrac{1}{2} 
\delta {\rm Z}_{ij}^{R *} \bold{P_R} \big) \tilde{\chi}_j^{0~\rm renormalized}, 
\eeq 
where the index $ j $ is summed over $ j \in \{1, 2, 3, 4\}$. The wavefunction 
renormalization counterterms $\delta Z_{ij}^{L/R}$ are determined using on-shell 
renormalization schemes \cite{Fritzsche:2002bi}. A comparison among different 
choices of the input masses can be found in Ref.\cite{Chatterjee:2011wc}.
The relevant Lagrangian may be expressed as, 
\begin{align}
\mathcal{L} = \mathcal{L}_{\rm Born} + \mathcal{L}_{CT}.
\end{align}
where the ``tree-level" Lagrangian $\mathcal{L}_{\rm Born}$ is expressed in terms of 
renormalized fields, and $\mathcal{L}_{\text{CT}}$ includes contributions from the 
relevant counterterms. The counterterm Lagrangian $\mathcal{L}_{\text{CT}}$ is given by, 
\beq \label{eq:LCT}
-\mathcal{L}_{CT} \supset 
 \overline{\tilde{\chi}^0_1} \gamma^{\mu} (\delta \mathcal{N}_{11}^L \bold{P_L} +  \delta \mathcal{N}^R_{11} \bold{P_R})
 \tilde{\chi}^0_1 Z_{\mu}.
\eeq
where the values of $\delta \mathcal{N}^{L/R}_{11}$ are given by,
\begin{align}\label{CT1}
\delta \mathcal{N}^{R}_{11} = \dfrac{e}{4 c_W^3 s_W^2}\Bigg[& \sum_{i=1}^{4}{\big(\delta {Z}^{R *}_{1i}+\delta 
Z^{R}_{i1}\big)(N_{13} N_{i3}-N_{14} N_{i4})s_W c_W^2} - \bigg(\big(2 \delta s_W-(2 \delta Z_e+\delta 
Z_{ZZ})s_W\big)c_W^2  \nonumber\\
& -2 \delta s_W s_W^2\bigg)\bigg(|N_{13}|^2-|N_{14}|^2\bigg)\Bigg],
\end{align}
and 
\begin{align}\label{eq:CT2}
\delta \mathcal{N}^{L}_{11} = -\dfrac{e}{4 c_W^3 s_W^2}\Bigg[& \sum_{i=1}^{4}{\big(\delta {Z}^{L *}_{1i}+\delta 
Z^{L}_{i1}\big)(N_{13} N_{i3}-N_{14} N_{i4})s_W c_W^2} - \bigg(\big(2 \delta s_W-(2 \delta Z_e+\delta 
Z_{ZZ})s_W\big)c_W^2  \nonumber\\
& -2 \delta s_W s_W^2\bigg)\bigg(|N_{13}|^2-|N_{14}|^2\bigg)\Bigg].
\end{align}
In Eqs.(\ref{CT1}) and (\ref{eq:CT2}), $\delta s_W$ is the counterterm for $s_W$, $
\delta Z_e$ is the counterterm corresponding to the electric charge $e$, and the $\delta 
Z^{L/R}_{ij}$ correspond to the wavefunction renormalization counterterms, as 
described in Eq.(\ref{eq:RenChi}). Further, $\delta Z_{ZZ}$ is the wavefunction 
renormalization counterterm for the $Z$-boson, and is given by, 
\beq
\delta Z_{ZZ} = - \left[\widetilde{\operatorname{Re}}\right] \Sigma_{Z}^{'T}(M_Z^2),
\eeq 
where the square brackets denote that $\delta Z_{ZZ}$ is diagonal and 
$\widetilde{\mathrm{Re}}$ takes the real part of loop integrals without affecting 
the complex couplings, which may be involved. The details of the counterterms 
in Eqs.(\ref{CT1}) and (\ref{eq:CT2}) may be found in Refs.\cite{Fritzsche:2013fta, 
Bisal:2023fgb}. We have closely followed the convention of 
Refs.\cite{Fritzsche:2002bi, Fritzsche:2013fta}. 

Thus, at one-loop level, the radiative corrections to the $\tilde{\chi}_1^0-$nucleon 
interactions receive contributions from the ``tree-level" interactions, the loop 
contributions,  and the counterterm contributions. As already discussed, in the 
present context we will consider radiative corrections to the 
$\tilde{\chi}_1^0\tilde{\chi}_1^0 Z$ vertex and the respective vertex counterterms, 
which contribute substantially to the spin-dependent $\tilde{\chi}_1^0-$ nucleon 
interactions.

For both Wino-like and Higgsino-like $\tilde{\chi}_1^0$ scenarios, substantial contribution 
to the spin-dependent $\tilde{\chi}_1^0$-nucleon come from the electroweak box diagrams 
(involving the gauge bosons, charginos, neutralinos and quarks), as shown in 
Fig.\ref{fig:Twist-2} \cite{Hisano:2004pv}.  The relevant contribution to the effective 
interaction strength $d_q$, as described in Eq.(\ref{eq:Leffq}), $d_q^{\rm box}$ 
has been discussed in Appendix A following Refs.\cite{Hisano:2004pv,Hisano:2012wm}.

%%%%%%%%%%%%%%%%%%%%%%%%%%%%%%%%%%%%%%%%%%%%%%%%%%
%%%%%%%%%%%%%%%%%%%%%%%%%%%%
\section{Results}
\label{sec:results}
%\subsection{Implementation of the 

In this section, we discuss the numerical results. In particular, we evaluate the vertex 
corrections to the $\tilde{\chi}_1^0\tilde{\chi}_1^0Z$ and the respective counterterms 
in the on-shell renormalization scheme for some benchmark scenarios. Further, important 
contributions from the electroweak box diagrams have been included in estimating the 
spin-dependent cross-section following Refs.\cite{Hisano:2004pv,Hisano:2010fy,
Hisano:2011cs, Hisano:2012wm}.
There are a total of 292 diagrams that are evaluated, composed of six different types 
of topologies described in Fig.\ref{fig:Top1} in Appendix B. We have checked the UV finiteness of the 
radiative corrections (comprising of the triangle loop diagrams and the respective 
vertex counterterms). The results are discussed in the context of the Higgsino-like and the 
Wino-like  $\tilde{\chi}_1^0$ DM.

\subsubsection{Implementation} \label{sec:methodology}
We outline the steps followed to compute the $\tilde{\chi}_1^0\tilde{\chi}_1^0Z$ 
vertex corrections and the improved spin-dependent direct detection 
cross-section has been presented below. The numerical treatment is similar to the
previous work on spin-independent Higgsino scattering with nucleons \cite{Bisal:2023fgb}. 
The tools
$\mathtt{FeynArts}$-3.11~\cite{Hahn:2000kx, KUBLBECK1990165, Hahn:2001rv, Fritzsche:2013fta}, 
$\mathtt{FormCalc}$-9.10~\cite{Hahn:1998yk, Fritzsche:2013fta}, $
\mathtt{LoopTools}$-2.16~\cite{Hahn:1998yk}, $\mathtt{SARAH}$-4.14.5~\cite{Staub:2017jnp,
 Staub:2013tta, Staub:2015kfa}, $\mathtt{SPheno}$-4.0.4~\cite{Porod:2003um, Staub:2017jnp}, and 
 $\mathtt{micrOMEGAs}$-5.3.41~\cite{Belanger:2006is, Belanger:2008sj, Belanger:2013oya}
 are utilized at different stages of the computations:
%\footnote{For a recent study related to DM-nucleon NLO cross-section, see \cite{Harz:2023llw}.}
\begin{itemize}
    \item First, the Feynman diagrams for the one-loop contributions to the
     $\tilde{\chi}_1^0\tilde{\chi}_1^0Z$ vertex and the relevant counterterms were evaluated
      using $\mathtt{FeynArts}$ and $\mathtt{FormCalc}$ respectively. The relevant loop 
     integrals were expressed in terms of the Passarino-Veltman (PV) scalar integrals.   
     Subsequently, $\mathtt{Fortran}$ subroutines were prepared to evaluate the 
   the analytical expressions for the vertices and corresponding counterterms numerically. 
   
    \item Using the spectrum generator $\mathtt{SPheno}$, which employs the 
    $\mathtt{SARAH}$ generated model file for the MSSM, we generated the particle 
    spectrum and the relevant mixng matrices for the benchmark scenarios. Next, we 
    evaluated the numerical values for the vertex corrections and the relevant counterterms 
    using $\mathtt{LoopTools}$ for the benchmark scenarios. The counterterms are 
    determined using suitable variants of the on-shell renormalization scheme 
    \cite{Fritzsche:2002bi, Baro2009, Chatterjee:2011wc,Fritzsche:2013fta}.  
    At this stage, we checked that all UV-divergences cancel, yielding a finite 
    result for the $\tilde{\chi}_1^0\tilde{\chi}_1^0Z$ vertex corrections. 
    
   \item Finally, we updated the $\tilde{\chi}_1^0\tilde{\chi}_1^0Z$ vertex in 
   $\mathtt{micrOMEGAs}$ by including all the triangular topologies depicted in 
   Fig.\ref{fig:Top1}, and these are used to estimate the spin-dependent direct detection 
   cross-section for the benchmark scenarios, as described in Tables 
   \ref{tab:bp1} and \ref{tab:bp4}. 
	
	\item Additionally, we incorporate the one-loop diagram shown in Fig.\ref{fig:Twist-2} to 
	calculate the $W/Z$ box contribution to the neutalino-quark scattering process. While their 
	analytical expressions are available in the literature, they were not previously included in 
	$\mathtt{micrOMEGAs}$. 
\end{itemize}

\subsection{Higgsino-Like Neutralino}

The lightest neutralinos considered in this study are predominantly Higgsino-like, 
with a Higgsino fraction of $\gtrsim 99\%$, characterized by $|\mu| \ll |M_1|, M_2$. 
The Higgsino mass parameter $|\mu|$ is set to {0.3 TeV and 0.6 TeV for the 
two sets of benchmark points (BP1–BP4). For one benchmark point (BP5), $\mu$ is set 
to 1.05 TeV to satisfy the thermal relic abundance using tree-level annihilation 
cross-section. These benchmarks represent the compressed Higgsino spectrum.

In Eq.(\ref{eq:LchiZ}), it is apparent that the tree-level 
$\tilde{\chi}_1^0 \tilde{\chi}_1^0 Z$ vertex depends on ($|N_{13}|^2-|N_{14}|^2$), which is
sensitive to the sign and hierarchy of the parameters $M_1$ and $M_2$ for a fixed $\mu$, even
when $|\mu| \ll |M_1|, M_2$. 
Consequently, the percentage radiative corrections $\Delta \mathcal{N}^{L/R}$, are sensitive 
to the sign and the hierarchy of the $M_1$ and $M_2$ parameters. 
%Consequently, the relative 
%importance of the radiative corrections to the $\tilde{\chi}_1^0 \tilde{\chi}_1^0 Z$ vertex 
%can significantly vary with the choice of parameters $M_1$ and $M_2$. 
Thus, to demonstrate this, the Bino mass parameter $M_1$, and the Wino mass parameter $M_2$ 
are varied as ($\pm 5$, $\pm 4$) TeV and (4, 5) TeV, respectively. In BP5, in order
to achieve the thermal relic abundance (using the tree-level annihilation cross-section), $M_1$ 
is lowered to $-1.3$ TeV. The gluino mass 
parameter $M_3$ is fixed at 3 TeV to evade the LHC constraints on the mass of gluinos. 
Additionally, $\tan{\beta}$ is fixed at 10, and the pseudoscalar Higgs mass is set to 
$1.414$ TeV. The {trilinear soft-supersymmetry breaking term involving two stop 
squarks} was assigned as $T_t = -4$ TeV for BP1-BP3 and BP5, while for BP4, it was set to 
$T_t = 4$ TeV. The soft supersymmetry breaking parameters for BP1-BP3 are: 
$m_{\tilde{Q}_L} = 2.69$ TeV, $m_{\tilde{t}_R} = 2.06$ TeV. For BP4 and BP5, these 
parameters are: $m_{\tilde{Q}_L} = 3.50 $ TeV, $m_{\tilde{t}_R} = 4.03$ TeV. This variation 
was done to test 
the sensitivity of the radiative corrections to the trilinear coupling and the squark masses. 
Therefore, each choice of the benchmark point represents a different line on 
the plots in Fig.\ref{fig:HiggsinoCS}, each benchmark indicating a different characteristic 
of the Higgsino-like $\tilde{\chi}_1^0$ cross-section for the same value of $\mu$ 
parameter. Moreover,} all benchmark points successfully pass the collider bounds implemented 
in \texttt{SModelS} (version 2.3.0) 
\cite{Alguero:2021dig, Heisig:2018kfq, Dutta:2018ioj, Ambrogi:2017neo, Ambrogi:2018ujg,
Sjostrand:2014zea, Sjostrand:2006za, Beenakker:1996ch, Buckley:2013jua} as well as the direct 
detection constraints from the LUX-ZEPLIN {\cite{LZ:2022lsv,
LZCollaboration:2024lux}} and the PICO-60 \cite{PICO-60} experiments. 

The dominant contributions arise from topologies Figs.\ref{fig:Top1}(e) and \ref{fig:Top1}(f), 
which involve loops with top quarks and stop squarks. These loops contribute at a level of 
$\sim\mathcal{O}(10^{-2})$ to the total vertex factor, which is $\sim\mathcal{O}(10^{-2})$. 
Contributions from other diagrams within topologies (e) and (f) to 
the total vertex correction are relatively minor.  Specifically, the contributions from 
bottom quark-sbottom squark loops are of order $\sim\mathcal{O}(10^{-3})$, and diagrams 
involving other quark/squark flavors, apart from (s)top and (s)bottom, contribute at 
$\sim\mathcal{O}(10^{-6})$. The combined contributions from loops involving leptons and 
sleptons in topologies (e) and (f) are of order $\sim\mathcal{O}(10^{-4})$. 
In Fig.\ref{fig:Top1}(a), the topology with $W^{\pm}W^{\pm}\tilde{\chi}^{\pm}_i$ contributes 
subdominantly, with a value of $\sim\mathcal{O}(10^{-5})$. While the loop involving 
$W^{\pm}\tilde{\chi}^{\pm}_i\tilde{\chi}^{\pm}_j$ in Fig.\ref{fig:Top1}(b) 
contributes even less, approximately $\mathcal{O}(10^{-6})$. Topologies shown in 
Figs.\ref{fig:Top1}(c) and \ref{fig:Top1}(d) have very small contributions of order 
$\sim\mathcal{O}(10^{-7})$. 
Finally, the contribution from diagrams involving both neutral and charged Higgs 
{bosons}, as well as electroweakinos in topologies (e) and (f), are negligibly 
small. For topology (e), they contribute at $\sim\mathcal{O}(10^{-7})$, and for topology (f), 
the contribution is also of the order $\sim\mathcal{O}(10^{-7})$.
      
\subsubsection{Benchmark Scenarios} 
In this subsection, the benchmark scenarios have been discussed. 
The benchmark points have been described in Table \ref{tab:bp1}. There are a total of nine
benchmark points (BP's) that we have discussed {in this subsection.}

\begin{table}[h!]
\begin{center}
 \begin{tabular}{|c|c|c|c|c|c|c|c|c|c|c|}
 \hline
 Parameters &  BP1a & BP1b  & BP2a & BP2b & BP3a & BP3b & BP4a & BP4b & BP5 \\
%  \hline
  \noalign{\hrule height 0.5mm} % Bold line
$\mu$ (GeV) & 300  & 600 &  {-300} & {-600} & 300 & 600 & 300 & 600 & 1050 \\
  \hline
 $M_1$ (GeV) &  -5000 & -5000 &  -4000 & -4000 & -4000 & -4000 & -5000 & -5000 & -1300 \\
\hline
 $M_2$ (GeV) &  4000 & 4000 &  5000 & 5000 & 5000 & 5000 & 4000 & 4000 & 4000 \\
%\hline
\noalign{\hrule height 0.5mm} % Bold line
%\hline
 $m_{\tilde{\chi}^0_1}$ (GeV)  & 299.17 & 599.06  &  299.62 & 599.57 & 299.44 & 599.36 & 299.17 & 599.06 & -1048  \\
 \hline
 $m_{\tilde{\chi}^0_2}$ (GeV)  & -300.44  & -600.39 &  -300.4 & -600.4 & -300.29 & -600.24 & -300.44  & -600.39 & 1049\\
\hline
 $m_{\tilde{\chi}^0_3}$ (GeV)  & 4002  & 4002 &  -4000 & -4000 & -4000 & -4000 & 4002 & 4002 & -1303 \\
\hline
 $m_{\tilde{\chi}^0_4}$ (GeV)  & -5000  & -5000 & 5001 & 5001 & 5001 & 5001 & -5000 & -5000 & 4002 \\
\hline
 $m_{\tilde{\chi}^{\pm}_1}$ (GeV)  & 299.56 & 599.43 & 300.18 & 600.10 & 299.67 & 599.58 & 299.56 & 599.43 & 1049 \\
\hline
 $m_{\tilde{\chi}^{\pm}_2}$ (GeV)  & 4002 & 4002  &  5001 & 5001 & 5001 & 5001 & 4002 & 4002 & 4002 \\
 \hline
$m_{h_1}$  (GeV)  &   125.28 & 125.25  & 125.15 & 125.08 &  125.19 & 125.16 & 125.64 & 125.53 & 124.42 \\
 \hline
$m_{h_2}$  (GeV)  &  1354   & 1280 &  1463.5  & 1496 &  1393 & 1359 & 1656 & 1856 & 901.19 \\
% \hline
\noalign{\hrule height 0.5mm} % Bold line
% \hline
 HF  & 0.9997  & 0.9996 &  0.9998 & 0.9998 & 0.9998 & 0.9997 & 0.9997 & 0.9996 & 0.9884 \\
\hline
 $N_{11} (\times 10^{\textrm{-}3})$  & -$6.291$  & 5.956 &  -6.344 & -5.933 & -$7.756$ & -7.252 & -$6.291$ & 5.956 & 107.35 \\
\hline
 $N_{12} (\times 10^{\textrm{-}2})$  & -$1.679 $ & 1.827 &  -1.081 & -1.156 & -$1.322$ & -1.412 &-$1.679 $ & 1.827 & 1.00 \\
\hline
$N_{13}$ & 0.708 & -0.707 & -0.708 & -0.707 & 0.708 & 0.707 & 0.708 & -0.707 & 0.704 \\
\hline
$N_{14}$ & -0.706 & 0.707 & -0.7065 & -0.707 & -0.706 & -0.707 & -0.706 & 0.707 & 0.702\\
\hline
$|N_{13}|^2$-$|N_{14}|^2$ ($\times 10^{-3}$) & 2.277  & 1.279 & 3.493 & 0.868 & 1.537 & 0.868 & 
2.277 & 1.279 & 2.806 \\
\hline
\end{tabular} 
\\
\caption{\label{tab:bp1} The benchmark scenarios with a Higgsino-like $\tilde{\chi}^0_1$ 
have been tabulated here. HF stands for Higgsino fraction. The {trilinear 
soft-supersymmetry breaking term involving two stop squarks} is set as 
$T_t= \mathrm{-} 4$ TeV for the 
benchmarks BP1-BP3 and BP5, and $T_t= 4$ TeV for BP4. The soft supersymmetry breaking 
parameters for the left and right type squarks and sleptons for BP1-BP3 are as follows: 
$m_{\tilde{Q}_L}=2.69$ TeV, $m_{\tilde{t}_R} = 2.06$ TeV, $m_{\tilde{b}_R}=2.50$ TeV, 
$m_{\tilde{L}_L}=2.06$ TeV, and $m_{\tilde{e}_R}=2.06$ TeV. For BP4 and BP5, these 
parameters are as follows: $m_{\tilde{Q}_L}=3.50$ TeV, $m_{\tilde{t}_R} = 4.03$ TeV, 
and the rest are the same as BP1-BP3. The following input parameters have been fixed 
for all the benchmark scenarios: 
$\tan{\beta}=10,~ m_{A} = 1.414$ TeV. The fixed input gluino mass parameter, 
$M_3=3$ TeV. Mass of the $Z$ boson, $M_Z = 91.18$ GeV. For the benchmarks 
BP1-BP3, the third generation squark mass and mixing parameters remain fixed: the 
lightest stop mass $ {m}_{\tilde{t}_1} = 2.05$ TeV, the heaviest stop 
mass ${m}_{\tilde{t}_2} = 2.71$ TeV, the lightest sbottom mass $ {m}_{\tilde{b}_1} = 2.50$ TeV, the 
heaviest sbottom mass  ${m}_{\tilde{b}_2}= 2.69$ TeV. For BP4 and BP5, the lightest stop mass 
${m}_{\tilde{t}_1} = 3.49$ TeV, the heaviest stop mass ${m}_{\tilde{t}_2} = 4.05$ TeV, the lightest 
sbottom mass $ {m}_{\tilde{b}_1} = 2.50$ TeV, the heaviest sbottom mass  ${m}_{\tilde{b}_2}= 3.50$ 
TeV. For all the benchmarks, the lightest stau mass $ {m}_{\tilde{\tau}_1} = 2.05$ TeV, and the 
heaviest stau mass $ {m}_{\tilde{\tau}_2} = 2.08$ TeV. The charged Higgs boson mass $M_{H^{\pm}}= 
1.416$ TeV, the CP-even Higgs mixing angle $ \alpha = \sin^{-1}(-0.1)$.}
\end{center}
\end{table}

All the required input parameters like $\mu$, $M_1$, and $M_2$ are mentioned 
in the tables, along with neutralino and chargino tree-level masses (as tree-level 
masses are used to evaluate the loop vertex factors). The SM-like light 
Higgs mass is taken as $m_{h_1}=125 \pm 0.7$ GeV, and $m_{h_2}$ is the 
mass of the CP-even heavy Higgs. The $N_{1i}$'s are the neutralino mixing matrix 
entries mentioned in Eq.(\ref{mnd}). HF is the Higgsino fraction of the lightest neutralino 
$=|N_{13}|^2+|N_{14}|^2$. It is evident from Table \ref{tab:bp1} that the lightest 
neutralino that we have considered is almost purely Higgsino-like ($\gtrsim 98.8\%$). For 
BP1, BP2, and BP4, $|M_1|$ and $M_2$ have not been varied, only the relative sign 
between $M_1$, $M_2$, and $\mu$ have been varied, and $|M_1|>M_2$. For BP3, 
$|M_1|$ and $M_2$ values are interchanged, and $M_2>|M_1|$. Because of this difference, 
the renormalization scheme used for these two types of benchmarks is also different. 
The same is true for BP5 as well. The renormalization scheme used to evaluate 
the counterterms for benchmarks BP1, BP2, and BP4 is \texttt{CCN[4]}, whereas 
the scheme used for BP3 and BP5 is \texttt{CCN[3]}.

The benchmark BP5 (approximately) satisfies the cosmological constraints on the 
relic density of dark matter with $\Omega h^2 = 0.123$ at this point. This is expected 
as for the Higgsino-like lightest neutralino, the correct relic density is observed at 
$\sim$1 TeV. All the benchmark points pass the \texttt{SModelS} tests as well as the 
spin-independent direct detection bounds of the LUX-ZEPLIN experiment
{(LZ) \cite{LZCollaboration:2024lux, LZ:2022lsv}, with benchmarks BP1 and 
BP3-BP5 lying in the 2$\sigma$ uncertainty band of the most recent LZ results. The 
spin-independent cross-sections (evaluated using the relevant tree-level 
vertices involving $\tilde{\chi}_1^0 $ in \texttt{micrOMEGAs}-5.3.41) are presented 
in Table \ref{tab:result} ($\sigma_{SI}^p$).} The spin-dependent bounds are less stringent, and all the points 
lie well below the spin-dependent direct detection bounds of PICO-60 \cite{PICO-60} 
(proton) and {LZ \cite{LZCollaboration:2024lux, LZ:2022lsv}} (proton and 
neutron). 

\subsubsection{Numerical Results and Discussion}
\begin{table}[h!]
\begin{center}
\begin{tabular}{|c|c|c|c|c|c|c|c|}
 \hline
\multirow{3}{*}{BP}  &  &    $ \Delta\mathcal{N}^{L/R}$ (\%)   & $W/Z$ box (\%) &   $\sigma_{\rm SD}^p$ [pb] & $\sigma_{\rm SD}^n$ [pb] &   \\
  & $\mathcal{N}^{L}$ (-$\mathcal{N}^{R}$) & Total   & $a_p$($a_n$) [GeV$^{-2}$] &  ($\Delta \sigma_{\rm SD}^p$ \%) &  ($\Delta \sigma_{\rm SD}^n$ \%) & $\sigma_{\rm SI}^p$ [pb] \\
  & &  (Loop, CT)  & & & & \\
 \hline
 \multirow{2}{*}{BP1a} &  $\textrm{-}3.72 \times 10^{\textrm{-}4}$ & -55.04  & 12.18 (-16.72) &  
 $6.37 \times 10^{\textrm{-}8}$ & $1.19 \times 10^{\textrm{-}8}$ & 2.79 $\times 10^{\textrm{-}11}$ \\
 &  & (5644, -5699) & -1.490$\times 10^{\textrm{-}9}$(-1.789$\times 10^{\textrm{-}9}$) & (-67.4) & (-92.0)& \\
 \hline
 \multirow{2}{*}{BP1b} & $\textrm{-}2.94 \times 10^{\textrm{-}4}$ & -36.73 & 12.52 (-17.22) &  $3.55 \times 10^{\textrm{-}8}$ & $1.00 \times 10^{\textrm{-}8}$ & 3.57$\times 10^{\textrm{-}11}$ \\
 &   & (10050, -10087) & -8.604$\times 10^{\textrm{-}10}$(-1.035$\times 10^{\textrm{-}9}$) & {(-42.6)} & {(-83.8)} & \\
 \hline
 \multirow{2}{*}{BP2a} &  $\textrm{-}1.66 \times 10^{\textrm{-}4}$ &  -70.23 & 18.02 (-24.74) &  {$2.03 \times 10^{\textrm{-}8}$} & $1.17 \times 10^{\textrm{-}10}$ & 4.84$\times 10^{\textrm{-}12}$ \\
   &  & (8389, -8459) & -1.489$\times 10^{\textrm{-}9}$(-1.787$\times 10^{\textrm{-}9}$) & (-77.2) 
   & (-99.7) & \\
 \hline
 \multirow{2}{*}{BP2b} &  $\textrm{-}2.00 \times 10^{\textrm{-}4}$ & -36.70 & 18.44 (-25.37) & 
 {$1.90 \times 10^{\textrm{-}8}$} & $3.13 \times 10^{\textrm{-}9}$ & 6.23$\times 10^{\textrm{-}12}$ \\
    &  & (14854, -14891) & -8.600$\times 10^{\textrm{-}10}$(-1.035$\times 10^{\textrm{-}9}$) & (-33.2) 
    & (-85.6) & \\
\hline
  \multirow{2}{*}{BP3a} &  $\textrm{-}1.39 \times 10^{\textrm{-}4}$ & -75.10 & 18.02 (-24.74) &  
  {$1.64 \times 10^{\textrm{-}8}$} & ${2.26 \times 10^{\textrm{-}13}}$ & 
  1.27$\times 10^{\textrm{-}11}$ \\
    &   & (8382, -8457) & -1.489$\times 10^{\textrm{-}9}$(-1.788$\times 10^{\textrm{-}9}$) &  (-81.6) 
    & (-100.0) & \\
 \hline
  \multirow{2}{*}{BP3b} & $\textrm{-}1.73 \times 10^{\textrm{-}4}$ &  -45.11  & 18.44 (-25.37) & 
  {$1.53 \times 10^{\textrm{-}8}$} & $1.90 \times 10^{\textrm{-}9}$ & 1.64$\times 10^{\textrm{-}11}$ \\
     &   & (14855, -14900) & -8.601$\times 10^{\textrm{-}10}$(-1.035$\times 10^{\textrm{-}9}$) & 
     (-46.3) & (-91.3) & \\
\hline
  \multirow{2}{*}{BP4a} & $\textrm{-}1.04 \times 10^{\textrm{-}3}$ &  25.31  & 12.18 (-16.72) & 
  {$3.69 \times 10^{\textrm{-}7}$} & $1.76 \times 10^{\textrm{-}7}$ & 2.67$\times 10^{\textrm{-}11}$ \\
   &  & (6059, -6034) &  -1.490$\times 10^{\textrm{-}9}$(-1.789$\times 10^{\textrm{-}9}$) & (89.0) 
   & (17.9) &  \\
 \hline
  \multirow{2}{*}{BP4b} &   $\textrm{-}6.30 \times 10^{\textrm{-}4}$ & 35.47 & 12.52 (-17.22) & 
  {$1.35 \times 10^{\textrm{-}7}$} & $6.61 \times 10^{\textrm{-}8}$ & 3.38$\times 10^{\textrm{-}11}$ \\
 &   & (10798, -10762) & -8.607$\times 10^{\textrm{-}10}$(-1.035$\times 10^{\textrm{-}9}$) &  
 (119.0) & (39.8) & \\
  \hline
  \multirow{2}{*}{BP5} &   $\textrm{-}1.13 \times 10^{\textrm{-}3}$ & 10.41 & 3.48 (-4.79) & 
  {$3.86 \times 10^{\textrm{-}7}$} & $2.54 \times 10^{\textrm{-}7}$ & 1.74$\times 10^{\textrm{-}10}$ \\
 &   & (4802, -4792) &   -5.247$\times 10^{\textrm{-}10}$(-6.319$\times 10^{\textrm{-}10}$) &  (29.7) & (11.6) &  \\
  \hline  
%  \hline  
\end{tabular}
\caption{\label{tab:result}  BP refers to the different benchmark points. 
$\mathcal{N}^{L}$/$\mathcal{N}^{R}$ are the tree level 
$\tilde{\chi}^0_1\tilde{\chi}^0_1 Z$ vertex factors. $ \Delta\mathcal{N}^{L/R}$ are 
the percentage corrections to the tree-level vertex factors. The loop and counterterm 
contributions are separately given in ``Loop" and ``CT", respectively. The total contributions 
from the loops and the counterterms are labelled as ``Total";   ``$W/Z$ box" refers to the 
percentage contribution 
to $a_p$($a_n$) from the spin-dependent (SD) component of the $W$ and $Z$ boson box 
diagrams that have been evaluated, as well as the magnitude of the box contribution 
to $a_p$($a_n$). The magnitude of the corrected proton (neutron) cross-sections is 
given as $\sigma_{\rm SD}^p$ ($\sigma_{\rm SD}^n$), and the percentage correction in 
the spin-dependent proton (neutron) cross-section is given as $\Delta \sigma_{\rm SD}^p$ 
($\Delta \sigma_{\rm SD}^n$). All the benchmarks pass the \texttt{SModelS} checks.}
\end{center}
\end{table}
In Table \ref{tab:result}, $\mathcal{N}^{L}$/$\mathcal{N}^{R}$ are the tree level vertex 
factors, as mentioned in Eq.(\ref{eq:LchiZ}) for the $\tilde{\chi}^0_1 \tilde{\chi}^0_1 Z$ 
vertex, and $\mathcal{N}^{L}=-\mathcal{N}^{R}$ for this vertex. 
$ \Delta\mathcal{N}^{L/R}=\dfrac{\mathcal{N}_{\mathrm{corr.}}^{L/R}-\mathcal{N}_{\mathrm{tree}}^{L/R}}{\mathcal{N}_{\mathrm{tree}}^{L/R}}\times 100$  (\%) are 
the percentage corrections to the tree-level vertex factors. The percentage corrections 
to the proton and neutron cross-sections are evaluated as $\Delta \sigma_{\rm SD}^{p/n}
=\dfrac{\sigma_{\rm SD~\mathrm{corr.}}^{p/n}-\sigma_{\rm SD~\mathrm{tree}}^{p/n}}
{\sigma_{\rm SD~\mathrm{tree}}^{p/n}}\times100$ (\%). 
``Loop" refers to the percentage contribution to the tree-level $\tilde{\chi}^0_1 \tilde{\chi}^0_1 Z$ 
vertex from all the loop diagrams evaluated as Loop = $\frac{\mathcal{N}^{L/R}_{\mathrm{Loop}}}
{\mathcal{N}^{L/R}_{\mathrm{tree}}} \times 100 \%$, and ``CT" refers to the 
counter term contribution to the tree-level vertex evaluated with the help of Eq.(\ref{CT1}) as CT 
= $\frac{\mathcal{N}^{L/R}_{\mathrm{CT}}}{\mathcal{N}^{L/R}_{\mathrm{tree}}} \times 100 
\%$. The box contribution is evaluated as $W/Z$ box (\%) $= \frac{a_{p/n(\textrm{box})}}{a_{p/n
(\textrm{tree})}}\times100 \%$.\footnote{Note that the sign of the tree-level 
amplitude for the $Z$ mediated diagram in \texttt{micrOMEGAs} has an opposite sign compared 
to the Ref.\cite{Hisano:2012wm}; therefore, the relative sign is adjusted for the box diagrams.}

\begin{figure}[h!]
	\centering
	\includegraphics[scale=0.3]{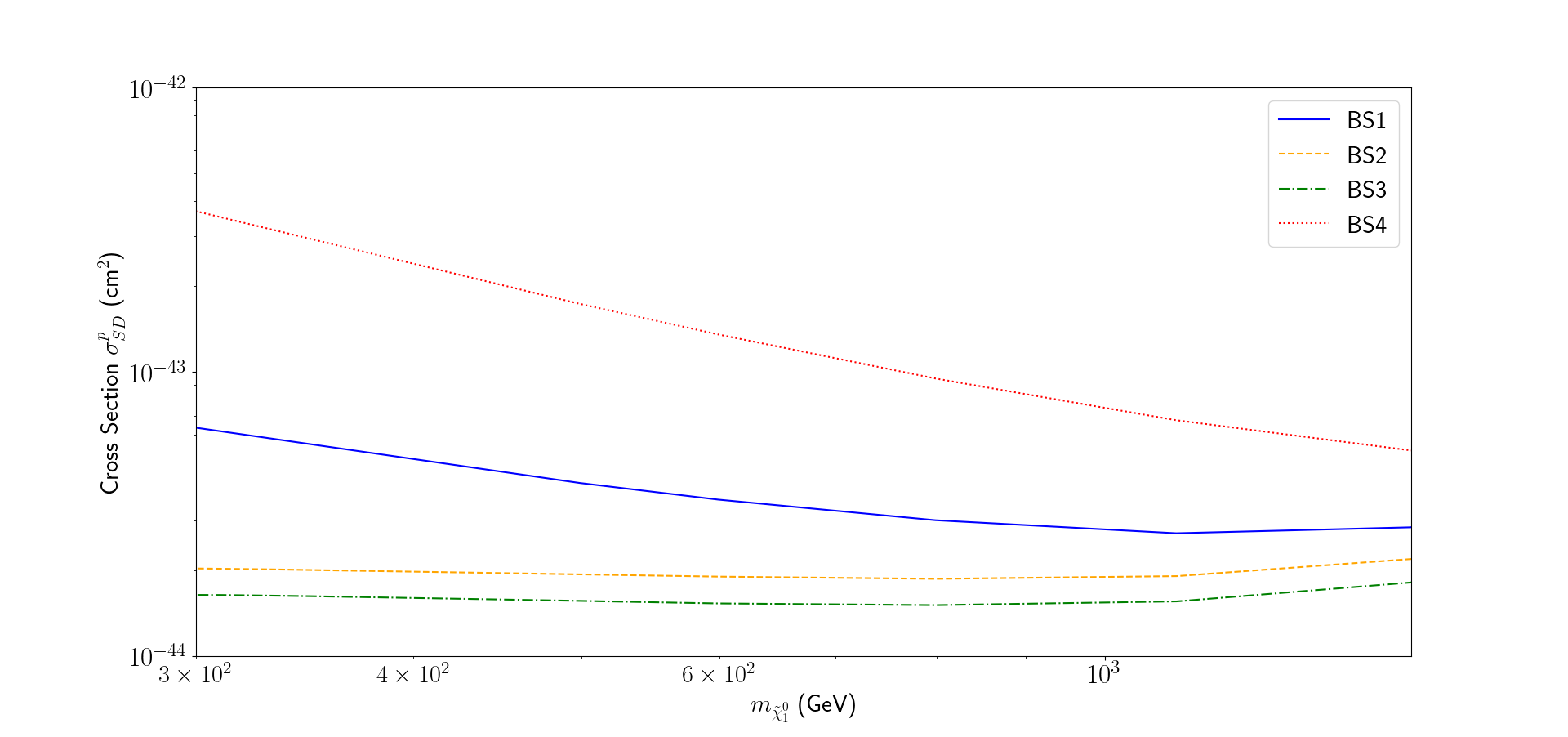} \\
	\quad \quad (a) \quad \\
	\includegraphics[scale=0.3]{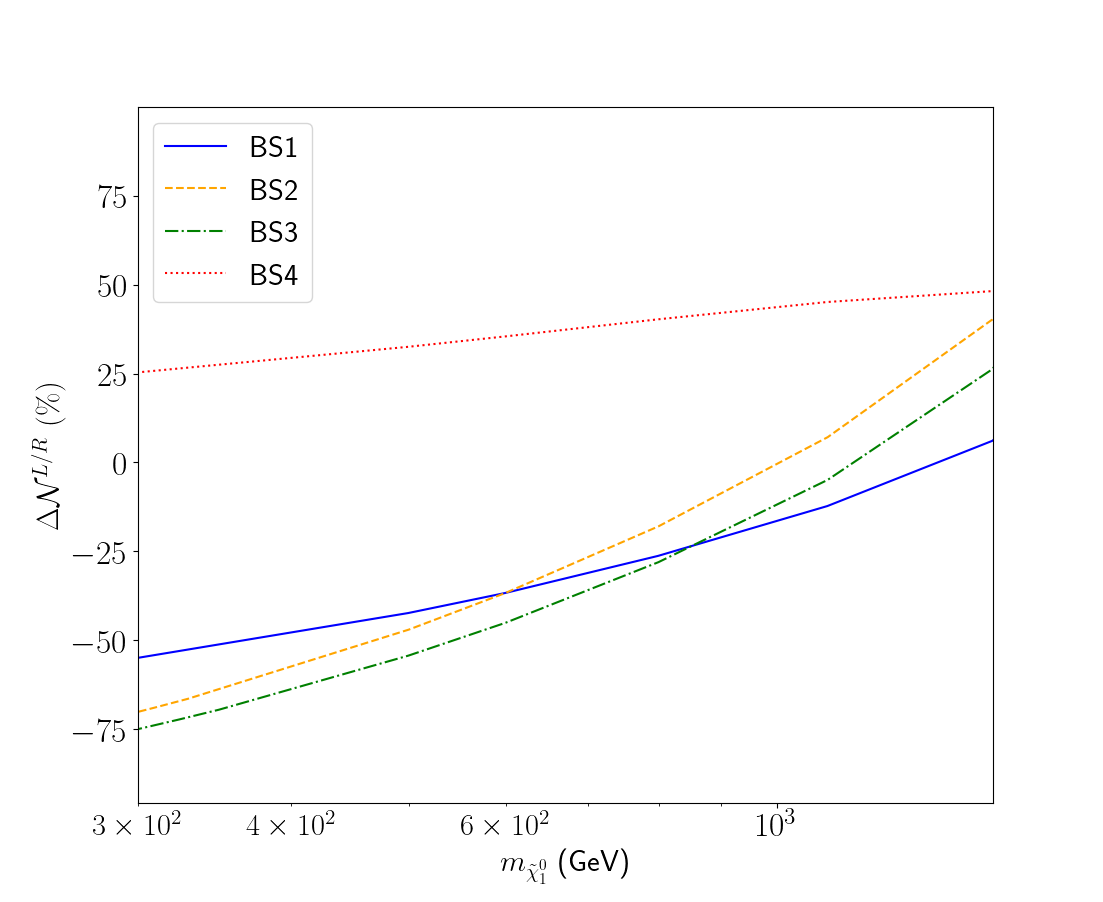}
	\includegraphics[scale=0.3]{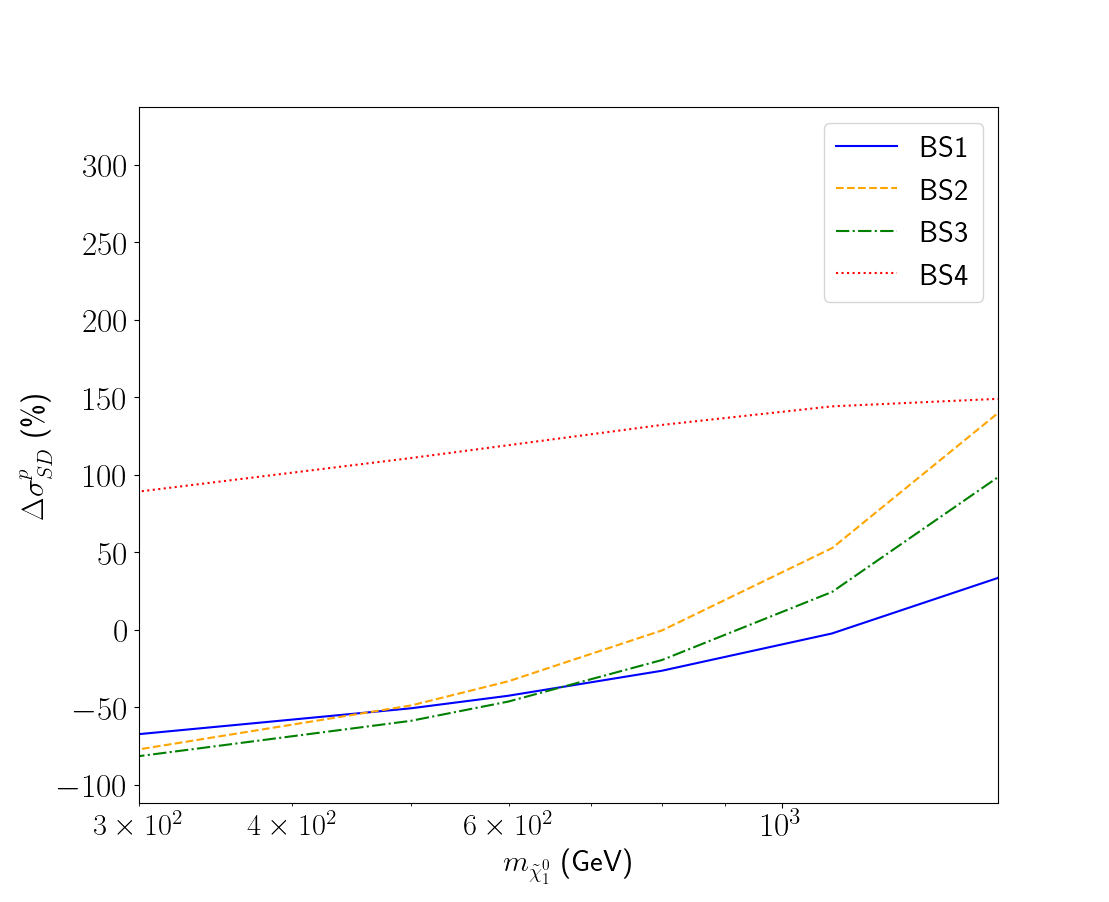} \\
	 (b) \hspace{8.2cm} (c)
	\caption{Panel (a) depicts the variation of the corrected spin-dependent 
	proton cross-section $\sigma_{SD}^p$ with $m_{\tilde{\chi}_1^0}$ for the benchmark 
	scenarios. 
	Panel (b) illustrates the $\tilde{\chi}_1^0\tilde{\chi}_1^0Z$ vertex corrections, 
	$\Delta \mathcal{N}^{L/R}$ $(\%)$, varying with respect to $m_{\tilde{\chi}_1^0}$. Panel 
	(c) shows the variation of the proton cross-section corrections, $\Delta \sigma_{SD}^p$ 
	(\%), with $m_{\tilde{\chi}_1^0}$. The labels ``BS1-BS4'' correspond to different 
	benchmark scenarios associated with the benchmark points BP1-BP4. 
	These scenarios represent continuous variations of the $\mu$ parameter, while all other 
	parameters remain fixed as specified in Table~\ref{tab:bp1}. The benchmark points 
	BP1-BP4 lie on these respective lines in the plot. All of the above plots represent the 
	Higgsino-like $\tilde{\chi}_1^0$ scenario.}
	\label{fig:HiggsinoCS}
\end{figure} 

For benchmark points BP1–BP3, we take the soft supersymmetry-breaking parameters
$m_{\tilde{Q}_L} = 2.69$ TeV and $m_{\tilde{t}_R} = 2.06$ TeV, while for BP4–BP5, 
$m_{\tilde{Q}_L}$ is increased to 3.50 TeV and $m_{\tilde{t}_R}$ to 4.03 TeV. This 
results in negative vertex and cross-section corrections for BP1–BP3, whereas positive 
corrections are observed for BP4–BP5. 
Significant corrections to the $\tilde{\chi}^0_1$-nucleon cross sections are found for 
Higgsino-like $\tilde{\chi}^0_1$ benchmark points. For BP1a, the total vertex correction 
is a notable -55.04\%. {Although the combined contribution from all loop 
diagrams exceeds the tree-level vertex factor by more than 56 times, 
sizeable cancellation between the loops and counterterms reduces the overall 
correction.} This gauge cancellation pattern is consistent across all benchmark points. 
The loop-corrected vertex factor is of the order $\mathcal{O}(10^{-4})$ for most 
benchmarks, except for BP4a and BP5, where it reaches $\mathcal{O}(10^{-3})$. 
The contribution from electroweak box diagrams ($W/Z$ box) is approximately 12\% for 
the proton ($a_p$) and -17\% for the neutron ($a_n$) in both BP1a and BP1b. After 
including the tree-level amplitude with all vertex loop diagrams, counterterms, and 
box diagram amplitudes, the total proton cross-section correction for BP1a is -67.4\%, 
while the neutron cross-section correction is -92\%. The spin-independent cross-section 
(proton) is of order $\mathcal{O}(10^{-11})$ for all benchmarks, except BP5, where 
a more mixed state results in a higher cross-section of order $\mathcal{O}(10^{-10})$. 
In BP1b, the total vertex correction is -36.73\%, with proton and neutron cross-section 
corrections of {-42.6\% and -83.8\%, respectively. For BP2a, 
the vertex correction is -70.23\%, with cross-section corrections of 
-77.2\% (proton) and -99.7\% (neutron). 
The $W/Z$ box contribution is $\sim$18\% for the proton and 
$\sim$-25\% for the neutron in BP2a and BP2b. In BP2b, the vertex 
correction is -36.70\%, while the cross-section 
corrections are -33.2\% for the proton and -85.6\% for the 
neutron.} For BP3a, the $\tilde{\chi}^0_1 \tilde{\chi}^0_1 Z$ vertex correction reaches 
-75.1\%. The $W/Z$ box contribution is around 18\% for the proton and -25\% for the 
neutron. The maximum correction for the neutron cross-section is -100\% for BP3a, with 
the proton cross-section correction of -81.6\%. In BP3b, the vertex correction 
is -45.11\%, with proton and neutron cross-section corrections of -46.3\% and 
-91.3\%, respectively. In BP4a, the vertex correction is 25.31\%, with proton and 
neutron cross-section corrections of 89\% and 17.9\%, respectively. The box 
contributions are approximately 12\% for the proton and -17\% for the neutron 
for both BP4 benchmarks. BP4b exhibits a vertex correction of 35.47\%, with 
a maximum proton cross-section correction of 119\% and a neutron cross-section 
correction of 39.8\%. For BP5, which is a more mixed state, the vertex correction 
is 10.41\%, with cross-section corrections of 29.7\% for the proton and 11.6\% 
for the neutron. The box contributions are relatively small, with proton and 
neutron amplitude corrections of 3.48\% and -4.79\%, respectively.

In the case of Higgsino-like $\tilde{\chi}_1^0$, the loops involving top quarks 
and stop squarks are the dominant contributors to the vertex corrections. Each 
of the two triangular loop diagrams—one involving two top quarks and a stop 
squark, and the other with two stop squarks and a top quark—contributes with 
a magnitude of $\gtrsim 0.025$ across all benchmark points. In contrast, the 
contributions from other diagrams, including those involving quarks, squarks, 
leptons, and sleptons, are comparatively small at 
$\lesssim \mathcal{O}(10^{-3})$.

The counterterm contributions are primarily driven by the $\delta Z^{L/R}_{12}$ 
and $\delta Z^{L/R}_{21}$ terms. While $\delta Z^{L/R}_{12}$ (and $\delta Z^{L/R}_{21}$) 
and $\delta Z^{L/R}_{11}$ are of comparable magnitude, $\sim \mathcal{O}(10^{-1})$, 
the $\delta Z^{L/R}_{11}$ term is suppressed by a factor of $(|N_{13}|^2 - |N_{14}|^2) 
\sim 10^{-3}$, as shown in Eqs.(\ref{CT1}) and (\ref{eq:CT2}). In contrast,  
$\delta Z^{L/R}_{12}$ and $\delta Z^{L/R}_{21}$ are not subject to this suppression, 
since $(N_{13}N_{23} - N_{14}N_{24}) \sim 1$. The terms $\delta Z^{L/R}_{i1}$ 
(and $\delta Z^{L/R}_{1i}$) for $i=3,4$ contribute $\mathcal{O}(10^{-4})$ for most 
benchmarks. Similarly, the counterterms $\delta Z_e$ and $\delta s_W$ 
contribute $\lesssim 10^{-2}$ but are also suppressed by the factor 
$(|N_{13}|^2 - |N_{14}|^2) \sim 10^{-3}$. Although $\delta Z_{ZZ}$ has a 
larger contribution, $\mathcal{O}(10^{-1})$, it is similarly suppressed by the 
same factor. This suppression pattern remains consistent across all benchmark 
points.

{
We have illustrated the results for the Higgsino-like $\tilde{\chi}_1^0$ in 
Fig.\ref{fig:HiggsinoCS}. The labels ``BS1-BS4'' in Fig.\ref{fig:HiggsinoCS} represent 
different benchmark scenarios corresponding to the benchmark points BP1--BP4. In these scenarios, 
the $\mu$ parameter, and thus, $m_{\tilde{\chi}_1^0}$ is varied, while all other parameters 
remain fixed as listed in Table~\ref{tab:bp1}. The benchmark points BP1--BP4 are located on 
their respective lines in the plot. In Fig.\ref{fig:HiggsinoCS}(a), shows the variation of the 
corrected spin-dependent proton cross-section ($\sigma^p_{SD}$) with $m_{\tilde{\chi}_1^0}$ in 
the Higgsino-like scenario.
In Fig.\ref{fig:HiggsinoCS}(b), the variation of the percentage vertex correction of the 
$\tilde{\chi}_1^0\tilde{\chi}_1^0Z$ vertex ($\Delta \mathcal{N}^{L/R}$) is illustrated with 
$m_{\tilde{\chi}_1^0}$ for different Higgsino-like benchmarks. Similarly, in 
Fig.\ref{fig:HiggsinoCS}(c), the variation of the percentage 
proton cross-section correction ($\Delta \sigma_{SD}^p$) is shown with $m_{\tilde{\chi}_1^0}$ 
for various Higgsino-like benchmarks. The various lines on the plot show the different 
benchmark scenarios where for the same value of the $m_{\tilde{\chi}_1^0}$, different values 
of the cross-section and vertex corrections are observed, emphasising the importance of the 
choice of benchmarks.
}

\subsection{Wino-like Neutralino}

{In the scenario where the lightest neutralino is almost purely Wino-like, the 
tree-level $\tilde{\chi}^0_1 \tilde{\chi}^0_1 Z$ vertex becomes negligible, and radiative 
corrections may become significant. Such a scenario is considered in the following 
subsection where $M_2 \ll |M_1|, |\mu|$, with a Wino fraction of $\gtrsim 99.9\%$. It is 
apparent from Eq.(\ref{eq:LchiZ}) that the tree-level 
$\tilde{\chi}_1^0 \tilde{\chi}_1^0 Z$ vertex depends on ($|N_{13}|^2-|N_{14}|^2$), which 
is sensitive to the sign and the hierarchy of $M_1$ and $\mu$ parameters for a fixed $M_2$. 
Consequently, the relative importance of the radiative corrections to the 
$\tilde{\chi}_1^0 \tilde{\chi}_1^0 Z$ vertex can significantly vary with the choice of
$M_1$ and $\mu$ parameters, even when $M_2 \ll |M_1|, |\mu|$. 
To demonstrate this, for each benchmark point, $M_2$ is varied between 1.5 TeV and 1.8 TeV, 
while $M_1$ and $\mu$ are varied between ($\pm 4$, $\pm 5$) TeV.} 
We have considered various combinations of $M_1$, $M_2$, and $\mu$ across different benchmark 
scenarios. In particular, BP6 and BP10 correspond to the hierarchy $|M_1| > |\mu|$, while 
BP7 to BP9 represent cases where $|M_1| = |\mu|$. For benchmarks with the same mass 
hierarchy between $|M_1|$ and $|\mu|$, we explore different relative signs between $M_1$,
$M_2$, and $\mu$ to capture a broader range of phenomenological effects. {$M_3$ 
is kept fixed at 3.5 TeV to evade the LHC collider constraints on the mass of gluinos}, 
with $\tan{\beta}$ set at 10 and the pseudoscalar Higgs mass held constant at 3.46 TeV. 
{Each benchmark scenario corresponds to a different line on the plot in 
Fig.\ref{fig:WinoCS}, further emphasising the significance of the benchmarks.} All 
benchmark points satisfy the collider constraints imposed by \texttt{SModelS} 
(version 2.3.0) \cite{Alguero:2021dig, Heisig:2018kfq, Dutta:2018ioj, Ambrogi:2017neo,
Ambrogi:2018ujg,Sjostrand:2014zea, Sjostrand:2006za, Beenakker:1996ch, Buckley:2013jua} as 
well as the direct detection bounds from {LZ 
\cite{LZCollaboration:2024lux, LZ:2022lsv}. The benchmark BP10 lies in the 2$\sigma$ 
uncertainty band for the spin-independent case. The spin-independent cross-sections 
(evaluated using the relevant tree-level vertices involving $\tilde{\chi}_1^0 $) are 
presented in Table \ref{tab:resultWino} ($\sigma_{SI}^p$).}

In the Wino case, the topology shown in Fig.\ref{fig:Top1}(a), which involves 
$W^{\pm}W^{\pm}\tilde{\chi}^{\pm}_i$, is one of the dominant contributors, with a 
contribution of $\sim\mathcal{O}(10^{-5})$ to the total vertex corrections (which are of 
order $\sim \mathcal{O}(10^{-5})$). The other dominant contribution comes from loops 
involving top quarks and stop squarks with both Figs.\ref{fig:Top1}(e) and 
\ref{fig:Top1}(f) contributing up to $\mathcal{O}(10^{-3})$. But there is a cancellation 
between these figures and the total contribution is of $\mathcal{O}(10^{-5})$.
The $W^{\pm}\tilde{\chi}^{\pm}_i\tilde{\chi}^{\pm}_j$ loop in Fig.\ref{fig:Top1}(b) 
contributes $\sim \mathcal{O}(10^{-6})$. The topologies in Fig.\ref{fig:Top1}(c) and 
\ref{fig:Top1}(d) have a much smaller contribution, of the order of 
$\sim\mathcal{O}(10^{-8})$. Other diagrams corresponding to topologies (e) and (f) 
contribute relatively less to the total vertex correction. The combined contributions 
from bottom-sbottom loops are of order $\sim \mathcal{O}(10^{-6})$, and contributions 
from all loops with other flavors of quarks/squarks (apart from (s)top 
and (s)bottom) are of the order $\mathcal{O}(10^{-8})$. Diagrams involving leptons and 
sleptons in topologies (e) and (f) contribute at $\sim \mathcal{O}(10^{-7})$. 
Diagrams involving both neutral and charged Higgses, as well as electroweakinos, in 
topology (e) contribute negligibly, at the level of $\sim \mathcal{O}(10^{-9})$. Topology 
(f) diagrams, however, contribute more significantly, at approximately 
$\mathcal{O}(10^{-6})$.

\subsubsection{Benchmark Scenarios} 
In this subsection, the Wino-like benchmark scenarios have been discussed. 
The benchmark points have been described in Table \ref{tab:bp4}. 

\begin{table}[h!]
\begin{center}
 \begin{tabular}{|c|c|c|c|c|c|c|c|}
 \hline
 Parameters &  					  	    BP6 & BP7 & BP8 & BP9 & BP10 \\
%  \hline
\noalign{\hrule height 0.5mm} % Bold line
 $M_2$ (GeV) & 					  	    1500 & 1500 & 1550  & 1600 & 1800 \\
  \hline
 $M_1$ (GeV) &  				  	   -5000 & 5000 & -5000 & -5000 & 5000 \\
 \hline
 $\mu$ (GeV)  &  				 	   -4000 & 5000 & 5000 & -5000 & 4000 \\
%\hline
\noalign{\hrule height 0.5mm} % Bold line
 $m_{\tilde{\chi}^0_1}$ (GeV)  & 	    1500 & 1499 & 1549 & 1600 & 1799 \\
 \hline
 $m_{\tilde{\chi}^0_2}$ (GeV)  & 	   -4000 & 4967 & -4973 & -4967 & -4000.5 \\
\hline
 $m_{\tilde{\chi}^0_3}$ (GeV)  &	    4001 & -5000 & 5001 & 5001 & 4001\\
\hline
 $m_{\tilde{\chi}^0_4}$ (GeV)  & 	   -5001 & 5034 & -5028 & -5034 & 5001\\
\hline
 $m_{\tilde{\chi}^{\pm}_1}$ (GeV)  &   1500 & 1499 & 1549 & 1600 & 1799\\
\hline
 $m_{\tilde{\chi}^{\pm}_2}$ (GeV)  &   4002 & 5002 & 5002 & 5001 & 4002\\
 \hline
$m_{h_1}$  (GeV)  &  				    124.39 & 125.70 & 125.79 & 124.18 & 125.53 \\
 \hline
$m_{h_2}$  (GeV)  &  				    4108 & 2743 & 2495  & 4205 & 2894\\
\noalign{\hrule height 0.5mm} % Bold line
% \hline
 WF  & 								    0.9995 & 0.9996 & 0.9996 & 0.9997 & 0.9991\\
\hline
 $N_{11} (\times 10^{\textrm{-}5})$&  2.742 & -10.83 & -5.943 & 1.426 & -21.96\\
\hline
 $N_{12}$  & 					       0.9997 & 0.9998 & -0.9998 & 0.9998 & 0.9996\\
\hline
$N_{13} (\times 10^{\textrm{-}2})$ &  2.238 & -1.809 & 1.823 & 1.724 & -2.617 \\
\hline
$N_{14} (\times 10^{\textrm{-}3})$ &  6.392 & 7.025 & -7.248 & 3.918 & 13.77 \\
\hline
$\Omega h^2$				 & 			0.092 & 0.092 & 0.098 & 0.110 & 0.121 \\
\hline
\end{tabular} 
\end{center}
\caption{\label{tab:bp4} The benchmark scenarios with a Wino-like $\tilde{\chi}^0_1$ have been 
tabulated above. WF stands for Wino fraction $= |N_{12}|^2$. $\Omega h^2$ is the thermal relic 
density of the lightest neutralino in the early universe. The {trilinear 
soft-supersymmetry breaking term involving two stop squarks} is set as $T_t= \mathrm{-} 4$ TeV 
for the given benchmarks. The pseudoscalar Higgs mass parameter $m_A = 3.46$ TeV. The soft 
supersymmetry breaking parameters for the left and right type squarks 
and sleptons are as follows: $m_{\tilde{Q}_L}=2.7$ TeV, $m_{\tilde{t}_R} = 3.05$ TeV, 
$m_{\tilde{b}_R}=3.50$ 
TeV, $m_{\tilde{L}_L}=3.05$ TeV, and $m_{\tilde{e}_R}=2.05$ TeV. The following input parameters 
have been fixed for all the benchmark scenarios: 
$\tan{\beta}=10$, the fixed input gluino mass parameter $M_3=3.5$ TeV, and the mass of the 
$Z$ boson $M_Z = 91.18$ GeV. The third generation squark mass and mixing 
parameters remain fixed: the lightest stop mass $ {m}_{\tilde{t}_1} = 2.65$ TeV, the heaviest 
stop mass ${m}_{\tilde{t}_2} = 3.10$ TeV, the lightest sbottom mass $ {m}_{\tilde{b}_1} = 2.69$ 
TeV, and the heaviest sbottom mass  ${m}_{\tilde{b}_2}= 3.50$ TeV. For all the benchmarks, the 
lightest stau mass $ {m}_{\tilde{\tau}_1} = 2.06$ TeV, and the heaviest stau mass 
$ {m}_{\tilde{\tau}_2} = 3.05$ TeV. The charged Higgs boson mass $M_{H^{\pm}}= 3.465$ TeV, and
the CP-even Higgs mixing angle $ \alpha = \sin^{-1}(-0.1)$.}
\end{table}

{The relic density of the Wino-like $\tilde{\chi}^0_1$ is satisfied considering the
(co-)annihilation processes at the tree-level for $M_2 \sim$ 1.8 TeV for the benchmark point 
BP10.\footnote{As mentioned in footnote 4, including Sommerfeld enhancement, the 
thermal relic is achieved at $M_2 \sim 2.7-3$ TeV \cite{Hisano:2006nn, Beneke:2014hja,
Hisano:2004ds,Beneke:2019qaa}. 
%For a detailed estimation, see e.g., the Ref.\cite{Beneke:2014hja}. 
We have not considered this effect in the analysis of the 
Higgsino-like and Wino-like $\tilde{\chi}^0_1$. However, we have checked for a benchmark 
point with $M_2=$ 2.7 TeV (sfermion masses are increased to 3-4 TeV to ensure 
$\tilde{\chi}^0_1$ is the LSP, keeping {other parameters similar to BP10), that the 
vertex correction is $\sim$ 8 \% and the cross-section correction is $\sim$ 70 \%, which is 
dominated by the electroweak box diagrams. The results are sensitive to the choice of $M_1$ and
$\mu$. Thus, qualitatively, we observe a similar trend regarding relative importance of the 
radiative corrections to the spin-dependent direct detection of Wino-like DM with 
$m_{\tilde{\chi}^0_1} \simeq 2.7-3$ TeV.}}} 
At the tree-level, the mass splitting between $\tilde{\chi}^0_1$ and $\tilde{\chi}^{\pm}_1$, i.e., 
$\Delta m = m_{\tilde{\chi}^{\pm}_1}-m_{\tilde{\chi}^0_1}$ is very small, $\sim \mathcal{O}(0.1)$ 
MeV. The small $\Delta m$ leads to the $\tilde{\chi}^{\pm}_1$ not decaying to the 
$\tilde{\chi}^0_1$ and coannihilation happens for the $\tilde{\chi}^{\pm}_1$ as well as the 
$\tilde{\chi}^0_1$, thus the relic density is increased. This leads to the correct relic density
of $\Omega h^2 \simeq 0.12$ at a lower $M_2$ value of 1.8 TeV. However, when the on-shell masses 
for the charginos and neutralinos are taken as the physical mass, then we get an increased mass 
splitting of $\Delta m = 155$-160 MeV. Then, the $\tilde{\chi}^{\pm}_1$ is no longer stable and 
can decay to $\tilde{\chi}^{0}_1$ and an electron. Thus, the benchmarks are not ruled out by 
stable charged particle constraints. The tree-level as well as the on-shell mass splitting 
$\Delta m$ is large for the case of Higgsino-like $\tilde{\chi}^0_1$, $\Delta m =$ 0.05-8 GeV. 
Therefore, the $\tilde{\chi}^{\pm}_1$ is not stable for the tree-level or the on-shell mass. 
Thus, for the almost pure Higgsino-like $\tilde{\chi}^{0}_1$, the correct relic density is 
achieved at the standard value of $\sim$1 TeV.

%\newpage
\subsubsection{Numerical Results and Discussion}
 
In the Wino-like $\tilde{\chi}^0_1$ scenario, in order to {evade the LHC 
(\texttt{SModelS}) constraints, $M_2$ has been considered $\gtrsim 0.65$ TeV
\cite{ATLAS:2021moa}. To probe the relic density satisfying parameter region,} the Wino 
mass parameter is raised to 1.5 TeV and above. This leads to more mixing between the 
electroweakinos, resulting in a relatively large tree-level amplitude. Hence, we see a 
diminished relative contribution from the vertex loop diagrams and the counterterms as 
compared to the Higgsino-like scenario. The vertex corrections are still 
phenomenologically significant and can exceed 15\% of the tree-level vertex. However, 
the most significant contribution here comes from the electroweak box diagrams or the 
$W/Z$ box diagrams shown in Fig.\ref{fig:Twist-2}. It is important to note here that 
from Eq.(\ref{eq:dq}), it is evident that when $Y=0$, which is the case for a Wino-like 
$\tilde{\chi}^0_1$, the contribution from the $Z$ boson box diagram vanishes. Thus, 
only the $W$ boson box diagram is responsible for the $W/Z$ box contribution in the 
cross-section corrections in the Wino-like $\tilde{\chi}^0_1$ scenario.\footnote{
Note that, as before, the sign of the tree-level amplitude for the $Z$ mediated diagram 
in \texttt{micrOMEGAs} has an opposite sign compared to the Ref.\cite{Hisano:2012wm}; 
therefore, the relative sign is adjusted for the box diagrams.}

\begin{table}[h]
\begin{center}
\begin{tabular}{|c|c|c|c|c|c|c|}
 \hline
\multirow{3}{*}{BP}  &  &    $ \Delta\mathcal{N}^{L/R}$ (\%)   & $W/Z$ box (\%) &   $\sigma_{SD}^p$ [pb] & $\sigma_{SD}^n$ [pb] & $\sigma_{SI}^p$ [pb]  \\
  & $\mathcal{N}^{L}$ (-$\mathcal{N}^{R}$)  & Total   & $a_p$($a_n$) [GeV$^{-2}$] &  ($\Delta \sigma_{SD}^p$ \%) &  ($\Delta \sigma_{SD}^n$ \%) &\\
  &  & (Loop, CT) & & & & \\
 \hline
  \multirow{2}{*}{BP6} &  $\textrm{-}1.72 \times 10^{\textrm{-}4}$ &  3.11 & 77.66 (-30.54) & 
   {$1.36 \times 10^{\textrm{-}8}$} & $1.34 \times 10^{\textrm{-}8}$ & 
   3.70$\times10^{\textrm{-}12}$ \\
  &   & (-0.24, 3.35) & -1.37$\times10^{\textrm{-}9}$(-1.37$\times10^{\textrm{-}9}$) & (231.3) & (-49.6) & \\
    \hline
 \multirow{2}{*}{BP7} &  $\textrm{-}1.15 \times 10^{\textrm{-}4}$ &  14.13 & 173.30 (-37.68) 
 &  {$7.41 \times 10^{\textrm{-}9}$} & $7.91 \times 10^{\textrm{-}9}$ & 1.93$\times10^{\textrm-11}$ \\
  &   & (13.80, 0.33) & -1.37$\times10^{\textrm-9}$(-1.37$\times10^{\textrm-9}$) & (799.6) 
  & (-54.6) & \\  
    \hline
 \multirow{2}{*}{BP8} &  $\textrm{-}1.13 \times 10^{\textrm{-}4}$ &  10.69 & 165.88 (-35.61) &
   {$6.92 \times 10^{\textrm{-}9}$} & $8.52 \times 10^{\textrm{-}9}$ & 2.04$\times10^{\textrm-11}$ \\
  &   & (13.30, -2.61) & -1.33$\times10^{\textrm-9}$(-1.33$\times10^{\textrm-9}$) & (717.35) 
  & (-53.6) & \\ 
\hline
 \multirow{2}{*}{BP9} &  $\textrm{-}1.18 \times 10^{\textrm{-}4}$ & {15.20} & 154.65 (-33.26) &  
 ${7.19 \times 10^{\textrm{-}9}}$ & ${9.98 \times 10^{\textrm{-}9}}$ & 9.56$\times10^{\textrm-13}$ \\
  &   & {(11.76, 3.45)} & -1.29$\times10^{\textrm-9}$(-1.29$\times10^{\textrm-9}$) & 
  ({721.2})  & ({-48.7}) & \\ 
  \hline 
 \multirow{2}{*}{BP10} &  $\textrm{-}1.90 \times 10^{\textrm{-}4}$ & 5.78  & 58.18 (-21.61) &  
 $1.44 \times 10^{\textrm{-}8}$ & $2.48 \times 10^{\textrm{-}8}$ & 6.58$\times 10^{\textrm{-}11}$ \\
 &  & (8.67, -2.89) &-1.16$\times10^{\textrm-9}$(-1.16$\times10^{\textrm-9}$)& (175.1) & (-34.6) & \\
 \hline
\end{tabular}
\caption{\label{tab:resultWino} BP refers to the different benchmark points. $\mathcal{N}^{L}$/$
\mathcal{N}^{R}$ are the tree level vertex factors. $ \Delta\mathcal{N}^{L/R}$ are the percentage 
corrections to the tree-level vertex factors. The loop and counterterm contributions are separately 
given in ``Loop" and ``CT", respectively, as in Table \ref{tab:result}. The total contributions 
from the loops and the counterterms are labelled as ``Total"; ``$W/Z$ box" refers to the percentage contribution to $a_p$($a_n$) from the spin-dependent (SD) component of the $W$ and $Z$ boson box 
diagrams that have been evaluated in Eq.(\ref{eq:an}), as well as the magnitude of the box 
contribution to $a_p$($a_n$). The magnitude of the corrected proton (neutron) cross-section is 
given as $\sigma_{SD}^p$ ($\sigma_{SD}^n$), the percentage correction in the spin-dependent 
proton (neutron) cross-section is given as $\Delta \sigma_{SD}^p$ ($\Delta \sigma_{SD}^n$). }
\end{center}
\end{table}

The results of the radiative corrections for the Wino-like neutralino are summarized in Table \ref{tab:resultWino}. The corrected vertex factors for all benchmark scenarios are of $\mathcal{O}(10^{-4})$. Notably, the loop and counterterm contributions remain small in these cases, and the large gauge cancellation observed in the Higgsino-like scenario is absent in the Wino-like case.

For BP6, the $\tilde{\chi}^0_1 \tilde{\chi}^0_1 Z$ vertex correction is 3.11\%, while the W boson box contribution is significant, resulting in a 77.66\% amplitude correction for the proton and -30.54\% for the neutron coming from the box diagram. Consequently, the total proton cross-section experiences a 231.3\% correction, whereas the total neutron cross-section undergoes a -49.6\% correction. The spin-independent proton cross-section is $\mathcal{O}(10^{-12})$, ranging from $\mathcal{O}(10^{-13})$ to $\mathcal{O}(10^{-11})$ across different benchmarks.
In the BP7 benchmark, the vertex correction is 14.13\%, with the largest box contribution seen in this case: 173.3\% for the proton and -37.68\% for the neutron. This leads to the most substantial cross-section correction, with a 799.6\% increase for the proton and a -54.6\% reduction for the neutron.
For BP8, the vertex correction is 10.69\%, and the W boson box correction reaches approximately 165\% for the proton and -35.6\% for the neutron. This results in a 717.35\% proton cross-section correction and -53.6\% for the neutron.
%In BP4, the vertex correction is 14.8\%, with the cross-section corrections being 761.1\% for the proton and -51.6\% for the neutron.
%For BP5, the vertex correction is 4.89\%, and the box corrections are 70.75\% for the proton and -27.34\% for the neutron, leading to cross-section corrections of 215\% for the proton and -43.8\% for the neutron.
In the BP9 benchmark, the vertex correction is {15.20\%}, and the box contributions 
are 154.65\% for the proton and -33.26\% for the neutron. The resulting cross-section corrections 
are {721.2\%} for the proton and {-48.7\%} for the neutron.
Finally, in BP10, the vertex correction is 5.78\%, with box contributions of 58.18\% for the 
proton and -26.61\% for the neutron, resulting in a proton cross-section correction of 175.1\% 
and a neutron correction of -34.6\%.
As illustrated in Table \ref{tab:resultWino}, the box contribution depends solely on the mass of 
the $\tilde{\chi}^0_1$, and shows no dependence on the parameters $M_1$ and $\mu$. This is 
evidenced by the identical magnitude of the box amplitude for the pair of benchmarks BP6 and BP7.
%, BP3 and BP4, and BP5 and BP6.

In the Wino-like $\tilde{\chi}^0_1$ scenario, as in the Higgsino-like case, the dominant loop 
contributions arise from the top quark and stop squark loops. Specifically, the triangular loop 
diagrams—one involving two top quarks and a stop squark, and the other involving two stop squarks 
and a top quark—both contribute with a magnitude of $\mathcal{O}(10^{-3})$. However, unlike the 
Higgsino case, these two triangle diagrams interfere destructively, resulting in an overall 
contribution of $\mathcal{O}(10^{-5})$. Despite this cancellation, these are still the dominant 
loop contributions, as the total loop contribution in the Wino-like $\tilde{\chi}^0_1$ scenario 
remains $\mathcal{O}(10^{-5})$. 
The next significant contribution comes from the W boson loops. The loops involving two W bosons 
and one chargino, along with those involving two charginos and one W boson, contribute at the 
order of $\mathcal{O}(10^{-5})$ to the vertex corrections. All other vertex diagrams involving 
particles such as leptons, sleptons, quarks, squarks, and Higgs bosons contribute subdominantly.

\begin{figure}[h!]
	\centering
	\includegraphics[scale=0.3]{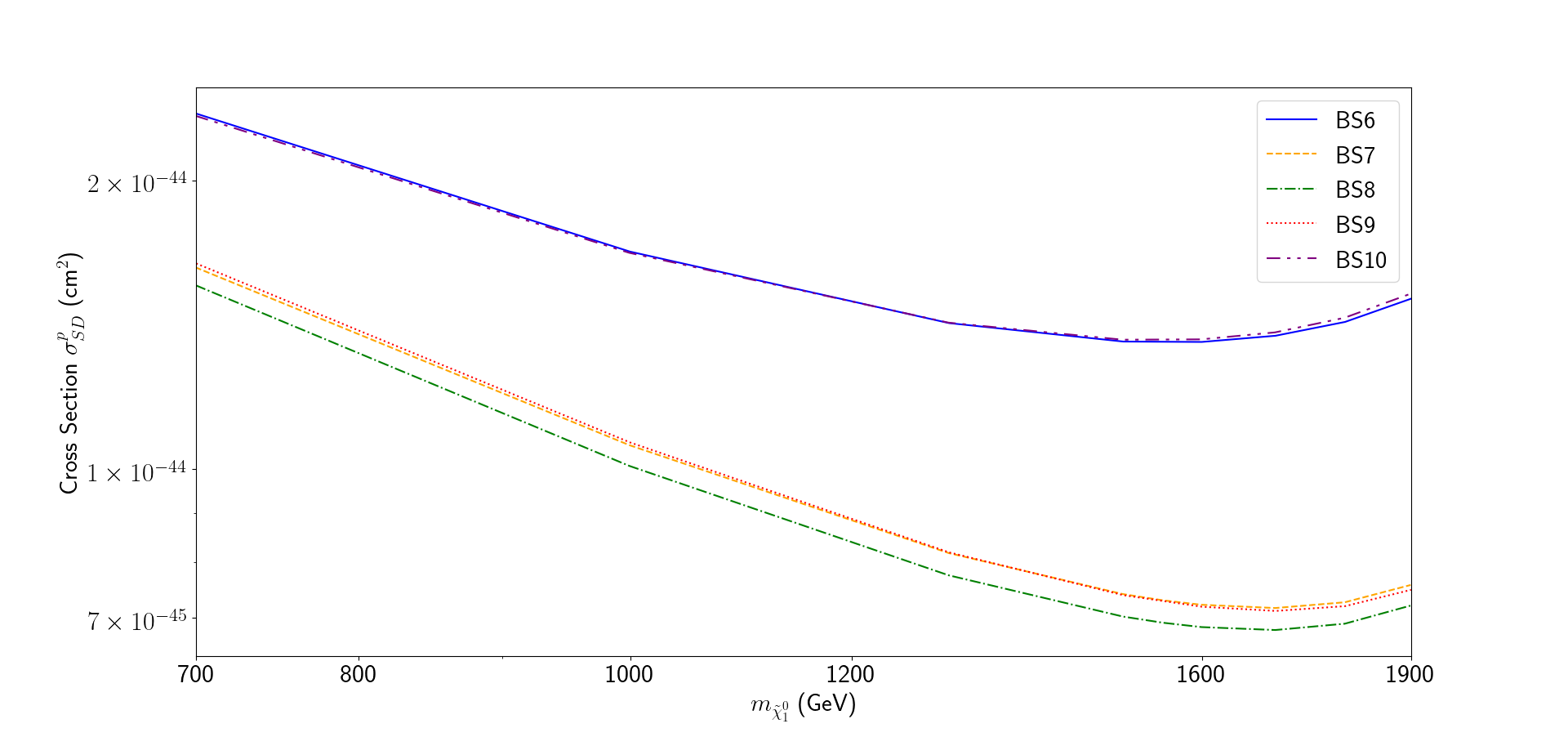} \\
		(a)	 \\
	\includegraphics[scale=0.3]{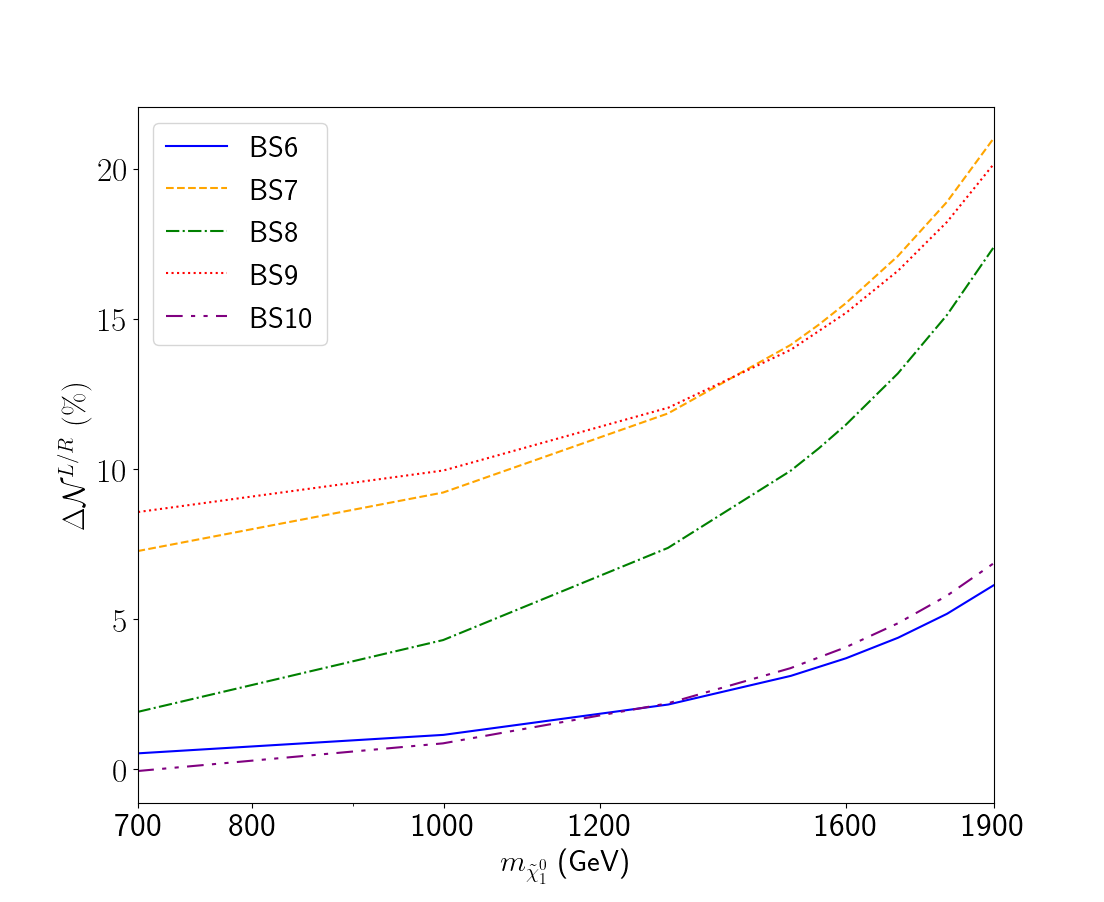} 
	\includegraphics[scale=0.3]{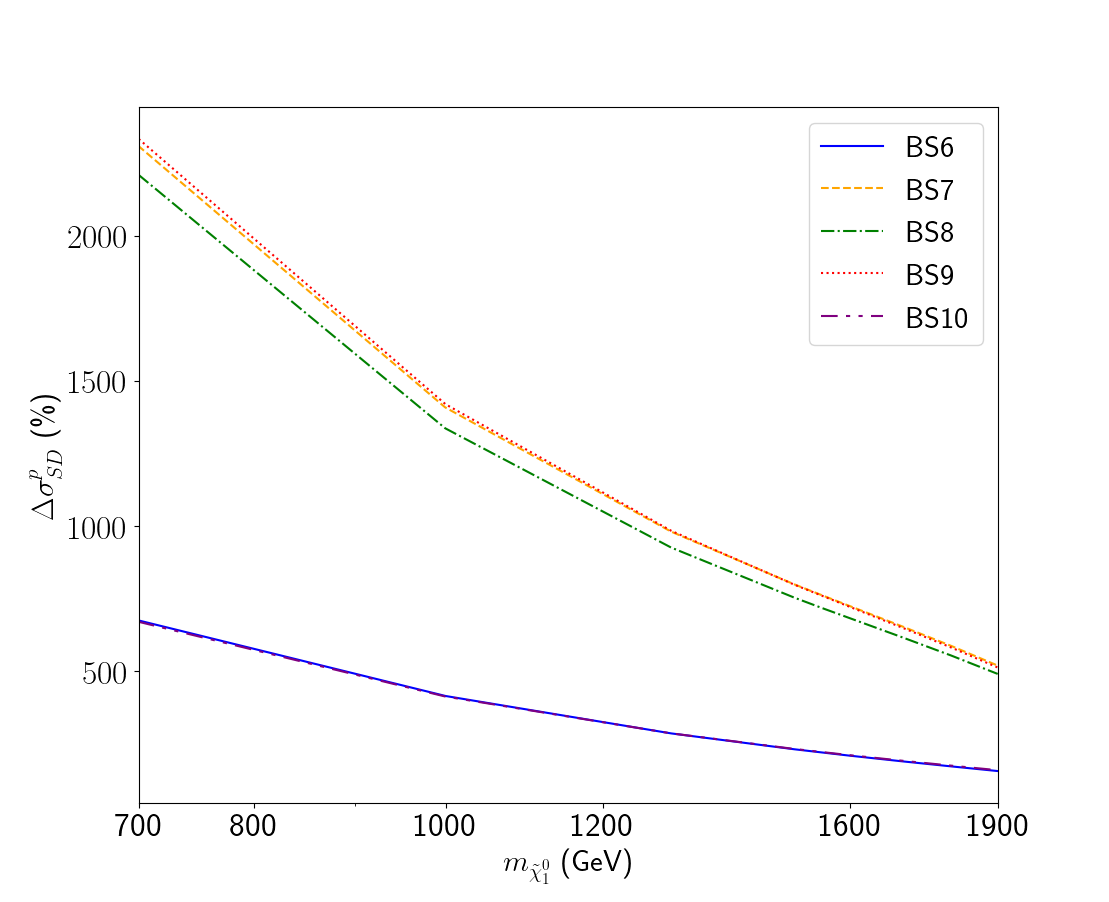} \\
			(b)	 \hspace{8.2cm}  (c)
	\caption{This figure represents the parameter plots for the Wino-like 
	$\tilde{\chi}_1^0$ scenario. Panel (a) depicts the variation of the corrected spin-dependent 
	proton cross-section $\sigma_{SD}^p$ with $m_{\tilde{\chi}_1^0}$. Panel (b) illustrates the 
	$\tilde{\chi}_1^0\tilde{\chi}_1^0Z$ vertex corrections, $\Delta \mathcal{N}^{L/R}$ $(\%)$, 
	varying with respect to $m_{\tilde{\chi}_1^0}$. Panel (c) shows the variation of the proton 
	cross-section corrections, $\Delta \sigma_{SD}^p$ (\%), with $m_{\tilde{\chi}_1^0}$.  
	Similar to Fig.\ref{fig:HiggsinoCS}, the ``BS6-BS10" labels mean the benchmark 
	scenarios corresponding to the benchmark points BP6-BP10, respectively, and only $M_2$ is 
	varied for those particular lines keeping all the other parameters same as given in Table 
	\ref{tab:bp4}.}
	\label{fig:WinoCS}
\end{figure} 

For the counterterm contributions, the $\delta Z^{L/R}_{11}$ term contributes up to 
$\mathcal{O}(10^{-1})$. However, it is suppressed by a factor of $(|N_{13}|^2 - |N_{14}|^2) 
\sim \mathcal{O}(10^{-4})$. On the other hand, the $\delta Z^{L/R}_{12}$ (and 
$\delta Z^{L/R}_{21}$) terms contribute up to $\mathcal{O}(10^{-3})$, but are multiplied by 
a factor of $(N_{13}N_{23} - N_{14}N_{24}) \sim \mathcal{O}(10^{-2})$ for benchmarks with 
$|M_1|>|\mu|$ (i.e., BP6 and BP10), which is less suppressed for a Higgsino-like 
$\tilde{\chi}_2^0$. Therefore, in such a scenario, both these terms end up contributing at 
a similar order. The terms involving $\delta Z^{L/R}_{13}$ and $\delta Z^{L/R}_{14}$ also 
contribute $\lesssim \mathcal{O}(10^{-3})$, but are usually subdominant compared to 
$\delta Z^{L/R}_{12}$. In the benchmarks with $|M_1|=|\mu|$, i.e., BP7-BP9, the factor 
$(N_{13}N_{23} - N_{14}N_{24}) \sim \mathcal{O}(10^{-3})$ and is smaller than the factor 
$(N_{13}N_{33} - N_{14}N_{34}) \sim \mathcal{O}(10^{-2})$ which multiplies the 
$\delta Z_{13}^{L/R}$ term. Therefore, in these benchmarks the $\delta Z^{L/R}_{11}$ 
and $\delta Z^{L/R}_{13}$ terms dominate, while $\delta Z^{L/R}_{12}$ and 
$\delta Z^{L/R}_{14}$ are suppressed. In the second term of the counterterm equations 
(\ref{CT1}) and (\ref{eq:CT2}), the $\delta Z_{ZZ}$ term is dominant, with a 
contribution of $\mathcal{O}(10^{-1})$. However, this term is also suppressed by a 
factor of $(|N_{13}|^2 - |N_{14}|^2) \sim \mathcal{O}(10^{-4})$. Finally, the terms 
proportional to $\delta s_W$ and $\delta Z_e$ contribute around $\mathcal{O}(10^{-2})$, 
but they are suppressed as $(|N_{13}|^2 - |N_{14}|^2)$ (see Eqs.(\ref{CT1}) and 
(\ref{eq:CT2})). This suppression pattern holds consistently for all benchmark points.

{
We have illustrated the results in Fig.\ref{fig:WinoCS} for the Wino-like $\tilde{\chi}_1^0$.
Fig.\ref{fig:WinoCS}(a) illustrates the variation of corrected proton cross-section 
($\sigma^p_{SD}$) with the ${\tilde{\chi}_1^0}$ mass. The ``BS6-BS10'' legends in 
Fig.\ref{fig:WinoCS} indicate that the benchmark points BP6-BP10 lie on their corresponding 
lines ``BS6-BS10'' in the plot. In these cases, only $M_2$ is varied, while all 
other parameters remain fixed as listed in Table~\ref{tab:bp4}.
Fig.\ref{fig:WinoCS}(b) and Fig.\ref{fig:WinoCS}(c) illustrate the variation of vertex 
correction and proton cross-section correction respectively, with ${\tilde{\chi}_1^0}$ mass, 
for different Wino-like benchmarks. The vertex corrections are increasing with mass of 
$\tilde{\chi}^0_1$ but the cross-section corrections are decreasing with increasing 
$m_{\tilde{\chi}^0_1}$ ($M_2$ parameter). This is due to the $W$ boson box diagram 
contributions, which are inversely related to the mass of $\tilde{\chi}^0_1$, and are the 
dominant contributors to the cross-section corrections (see Appendix A, especially 
Eqs.(\ref{eq:dq}) and (\ref{eq:mass})). The corrections vary smoothly for each benchmark 
scenario, which further demonstrates the significance of the choice of benchmark points.

\subsection{Comparison with direct detection experiments}

\begin{figure}[h!]
	\centering
	\includegraphics[scale=0.3]{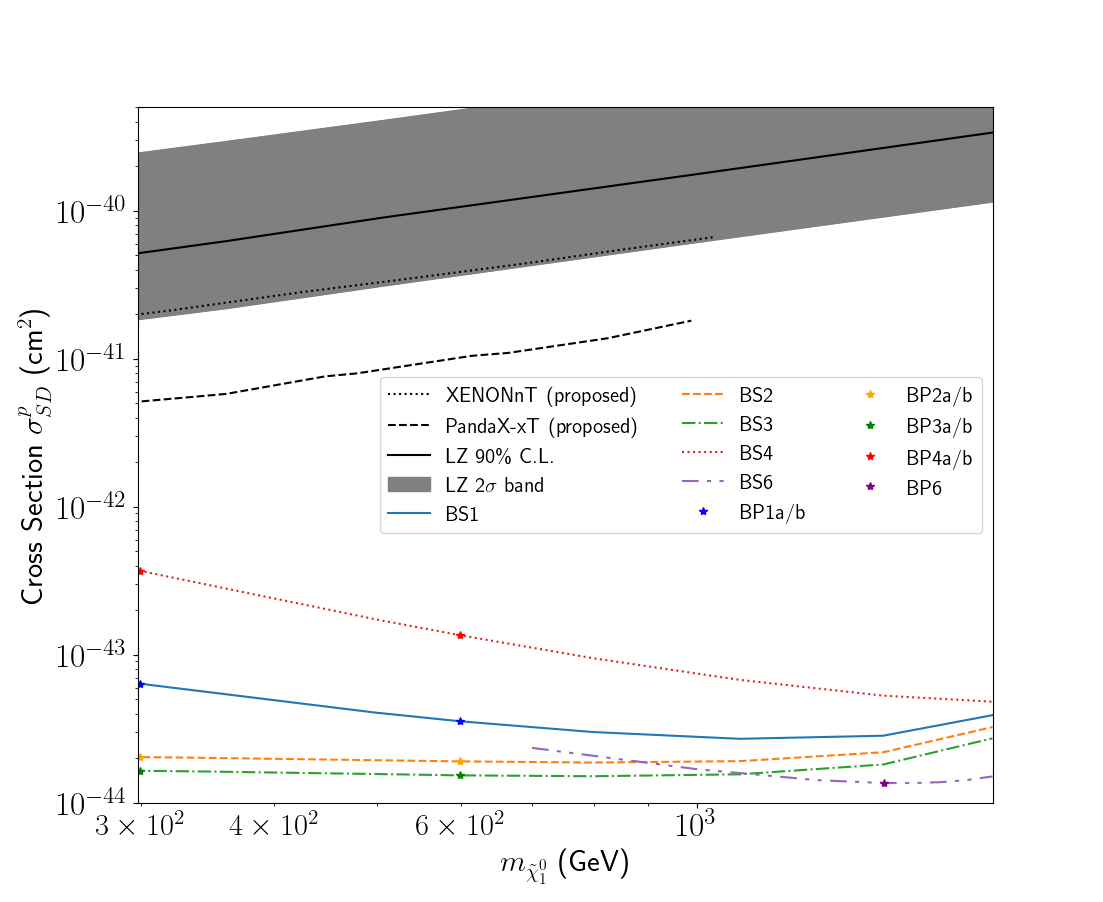} 
	\includegraphics[scale=0.3]{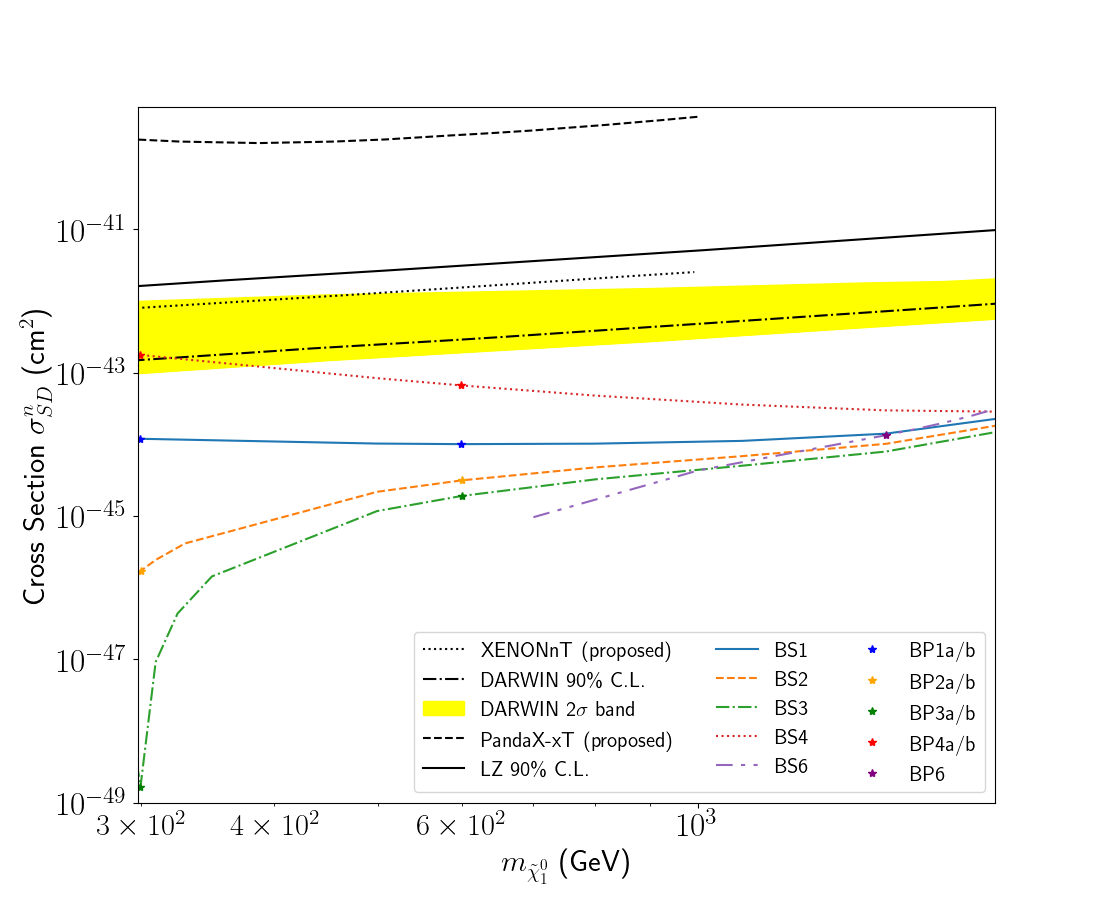} \\
			(a)	 \hspace{8.1cm}  (b)
	\caption{Panels (a) and (b) illustrate the comparison of the Higgsino-like 
	(and Wino-like) $\tilde{\chi}_1^0$ corrected cross-section comparison with current and 
	future direct detection bounds for proton and neutron cross-sections, respectively. The 
	``BS1-BS6" legend labels correspond to the benchmark scenarios related to the benchmark 
	points BP1-BP6, respectively. BP1-BP6 
	benchmark points (apart from BP5) are also plotted along with the benchmark scenarios, 
	for reference. The LZ bounds are taken from Ref.\cite{LZCollaboration:2024lux}, 
	PandaX-xT future experiment's proposed limits are taken from Ref.\cite{PANDA-X:2024dlo}, 
	the XENONnT proposed limits are from Ref.\cite{XENON:2020kmp} and the proposed limits 
	for the future direct detection experiment DARWIN are taken from 
	Ref.\cite{DARWIN:2016hyl}.}
	\label{fig:DDbounds}
\end{figure} 

The comparison of the corrected cross-section plots for the Higgsino-like and Wino-like 
$\tilde{\chi}_1^0$ with some prominent current and future direct detection experimental
bounds is shown in Fig.\ref{fig:DDbounds}. The results of this work are compared with the 
most recent bounds of LZ experiment \cite{LZCollaboration:2024lux}, the proposed limits 
of the ongoing XENONnT experiment \cite{XENON:2020kmp}, and proposed limits of the future 
experiments of PandaX-xT \cite{PANDA-X:2024dlo} and DARWIN \cite{DARWIN:2016hyl}. The 
cross-section bounds for the proton case is not mentioned in the DARWIN experiment's 
proposed plots, therefore it is not shown in Fig.\ref{fig:DDbounds}. All of the benchmark 
points lie below the current and future direct detection experimental limits apart from 
BP4a, which lies beyond the proposed 90\% C.L. of the DARWIN experiment, but still lies 
within the 2$\sigma$ uncertainty band. Thus, all the benchmarks are safe from experimental 
exclusion. The Wino-like $\tilde{\chi}_1^0$ cross-sections are much smaller than their 
Higgsino-like counterparts, and lie well below the experimental bounds. Thus, only one of 
the benchmark scenarios, i.e., BS6 (BP6) has been shown in the Figure, which has the 
maximum corrected cross-section in the Wino-like case. In this context, note that the
respective (tree-level) spin-independent cross-sections for all the benchmark points
except BP6 and BP9 can be probed, for example, in the proposed experiment DarkSide-20k
\cite{DarkSide-20k:2017zyg}. 
}

\section{Conclusion}
\label{sec:Conclusion}
To summarize, in this article we have explored the implications of important 
radiative corrections to the spin-dependent interaction of Higgsino-like and Wino-like 
neutralino $\tilde{\chi}_1^0$ DM with nuclei. In particular, we considered the radiative 
corrections to the $\tilde{\chi}_1^0\tilde{\chi}_1^0Z$ vertex, and evaluated the triangle 
loops and the respective vertex counterterms in appropriate  versions of the on-shell 
renormalization scheme. We then studied the implication of the vertex corrections to the 
spin-dependent direct detection of the $\tilde{\chi}_1^0$ DM. Further, we added 
contributions from important electroweak box diagrams, which have been 
previously studied in the literature, to estimate the impact of the radiative corrections 
to the spin-dependent $\tilde{\chi}_1^0$-nucleon direct detection cross-section. 

Specifically, for the benchmark scenarios with Higgsino-like $\tilde{\chi}_1^0$ 
substantial corrections to the tree-level $\tilde{\chi}^0_1\tilde{\chi}^0_1Z$ vertex are 
observed, with maximum corrections of approximately -75\% for the Higgsino-like case and 
15\% for the Wino-like case. At the cross-section level, these corrections can lead to 
changes of up to $\sim-$93.5\% and 57\% for the Higgsino- and Wino-like scenarios, 
respectively. Including the electroweak box diagram contributions, we observe 
cross-section corrections of approximately 119\% (-100\%) in the Higgsino-like case, 
and 800\% (-55\%) for the Wino-like case in the proton (neutron) cross-sections. 
These highlight the crucial impact of radiative corrections on the spin-dependent 
cross-sections for neutralino dark matter. The significant enhancement or suppression 
in interaction strength at the loop level, compared to the tree level, indicates that 
neglecting these corrections could lead to substantial under- or overestimations 
of detection rates in direct detection experiments. 

For  Higgsino-like $\tilde{\chi}_1^0$, the vertex corrections are sizable. 
Further, there is a large cancellation between the triangle loop contributions and the 
counterterm contributions evaluated using the on-shell renormalization scheme. 
Note that, in this case, the suppression of the $\tilde{\chi}^0_1 \tilde{\chi}^0_1 Z$ 
vertex is attributed to  a cancellation between the mixing matrix elements $N_{13}$ 
and $N_{14}$, which corresponds to the down-type and up-type Higgsino gauge 
eigenstate components to $\tilde{\chi}_1^0$. Thus the suppression of the tree-level 
vertex is generally also sensitive to the gaugino mass parameters. This has been 	
demonstrated by varying the gaugino mass parameters for the same Higgsino mass 
parameter $\mu$ in different benchmark scenarios. Among the triangle loops, largest 
contribution comes from the loops involving (s)top (s)quarks, thanks to the large 
Yukawa coupling. The overall vertex correction is sensitive to the stop mass and 
mixing parameters including the tri-linear soft-supersymmetry breaking term. 
This feature has been explored in different benchmark scenarios.  As for the 
counterterms, the largest contribution comes from the off-diagonal wavefunction 
renormalization constant $\delta Z_{12}$, which receives large contribution from 
the (s)top sector for the same reason as stated above. For  Wino-like $\tilde{\chi}_1^0$, 
the total effect of the vertex corrections is  generally much smaller than that of 
the electroweak box contributions, which dominate. Note that the loop contributions 
are also suppressed by the Wino-Higgsino mixing.
 
The benchmark scenarios demonstrate the significance of radiative corrections 
on the spin-dependent cross-sections for Higgsino-like (and also Wino-like) 
$\tilde{\chi}_1^0$. These corrections may enhance or deplete the spin-dependent 
cross-sections. While in the parameter region of interest, the estimated 
spin-dependent cross-section is much smaller than that of the experimental limits, 
it is nevertheless useful to estimate the spin-dependent $\tilde{\chi}_1^0$-nucleus 
scattering cross-sections accurately, which can be relevant in the light of future 
experiments. The present study, therefore, highlights the significance of radiative 
corrections to the spin-dependent $\tilde{\chi}_1^0$-nucleus interactions.

\section*{Acknowledgement}
The computations were supported in part by SAMKHYA, the High-Performance Computing 
Facility provided by the IoP, Bhubaneswar, India.
The authors acknowledge useful discussions with A. Pukhov. A.C. acknowledges the
hospitality at IoP, Bhubaneswar, during the meeting IMHEP-19 and IMHEP-22 which 
facilitated this work. A.C. and S.A.P. also acknowledge the hospitality at IoP, 
Bhubaneswar, during a visit. S.B. acknowledges the local hospitality at SNIoE, 
Greater Noida, during the meeting at WPAC-2023 where this work was put in motion.
S.B. and D.D. also acknowledge the local hospitality received at SNIoE, Greater Noida, 
during the final stage of the work.

% \newpage
\bigskip

\renewcommand{\theequation}{A.\arabic{equation}}
\setcounter{equation}{0}
\setcounter{section}{0}
\section*{Appendix A}
\label{sec:App-A}
%%%%%%%%%%
\subsection{$W$ and $Z$ Box Contribution}

\begin{figure}[h!]
	\centering
	\includegraphics[scale=0.5]{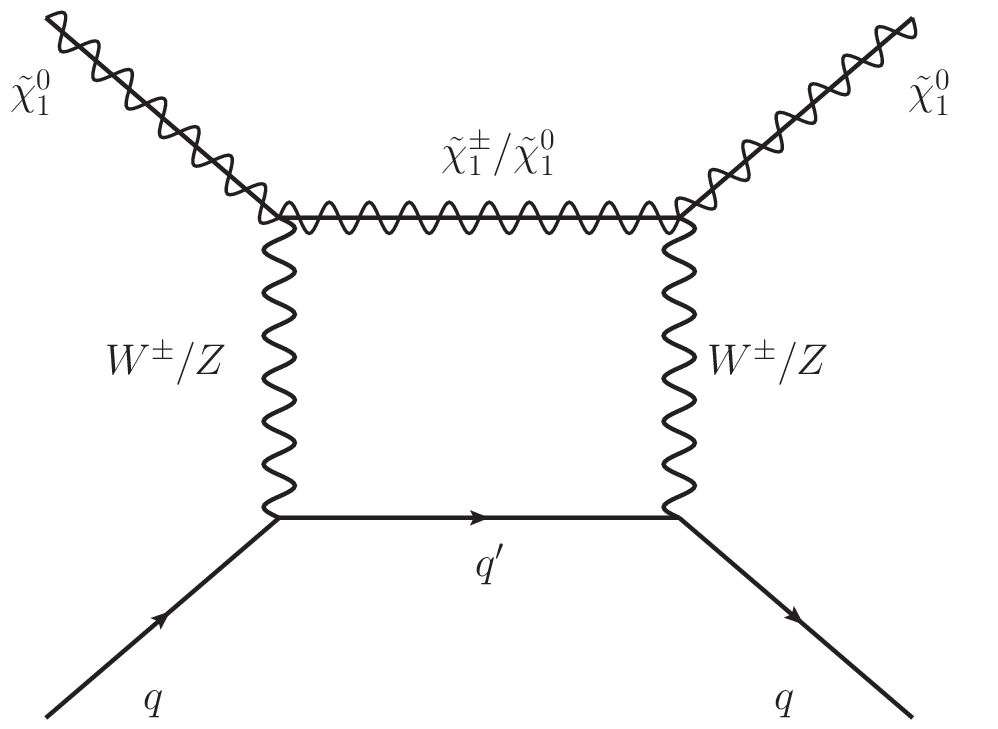}
	\caption{$W^{\pm}/Z$ box diagram for the neutralino-quark scattering process.}
	\label{fig:Twist-2}
\end{figure}

The $W/Z$ box diagrams also contribute towards the direct detection process of neutralino dark matter. 
The $W$ and $Z$ boson box diagrams as as shown in Fig.\ref{fig:Twist-2}, play a role in the enhancing 
(or reducing) the spin-dependent cross-section of the $\tilde{\chi}^0_1 \tilde{\chi}^0_1 Z$ vertex.
The coefficient of the amplitude of the neutralino-quark scattering via the $W$ and $Z$ box diagrams 
is given as \cite{Hisano:2011cs,Hisano:2012wm}
\begin{align}\label{eq:an}
a_N = \sum_{q=u,d,s}{d_q \Delta q_N}
\end{align}
(where $N=p,n$), and
\begin{align}
d_q = d_q^{\mathrm{tree}} + d_q^{\mathrm{box}},
\end{align}
where
\begin{align}\label{eq:dq}
d_q^{\mathrm{box}} = \dfrac{\mathrm{n}^2 - (4Y^2 +1)}{8} \dfrac{\alpha_2^2}{m_W^2} g_{AV}(w) + \dfrac{2((a^V_q)^2-(a^A_q)^2)Y^2}{\cos^4{\theta_W}} \dfrac{\alpha_2^2}{m_Z^2} g_{AV}(z).
\end{align}
\\
Here, $m_W$ and $m_Z$ are the masses of $W$ and $Z$ bosons respectively, and $\alpha_2^2 = g_2^2/4\pi$.
The coupling of quarks with $Z$ boson (vector and axial-vector) is given as 
\begin{align}
a^V_q = \dfrac{1}{2} T_{3q}-Q_q \sin^2{\theta_W}, \hspace{1cm}  a^A_q = -\dfrac{1}{2} T_{3q}.
\end{align}
Here, $T_{3q}$ and $Q_q$ represent the weak isospin and charge of quark q, respectively. Further, 
we parametrize $w = m_W^2/m_{\tilde{\chi}_1^0}^2$ and $z = m_Z^2/m_{\tilde{\chi}_1^0}^2$ in the 
above equations. The mass function $g_{AV}(x)$ is defined in Eq.(\ref{eq:mass}). Y is the hypercharge 
and n is the n-tuplet. We take n $=2$ and $Y^2=\frac{1}{4}$ for the Higgsino case (pure 
limit), and n $=3$ and $Y=0$ for the Wino case (pure limit). $T_{3u}=\frac{1}{2}$ and $T_{3d/s}=
-\frac{1}{2}$, meanwhile $Q_u=\frac{2}{3}$ and $Q_{d/s}=-\frac{1}{3}$. $\theta_W$ is the weak mixing 
angle. The experimental values of the quark contribution to the spin of the nucleons for the light 
quarks, the $\Delta q_N$'s, are taken to be \cite{Belanger:2008sj}
\begin{subequations}
\begin{align}
\Delta u_p &= 0.842, \hspace{1cm}  \Delta d_p = -0.427, \hspace{1cm} \Delta s_p = -0.085,
\\
\Delta d_n &= 0.842, \hspace{1cm}  \Delta u_n = -0.427, \hspace{1cm} \Delta s_n = -0.085.
\end{align}
\end{subequations}
These values are confirmed by the HERMES collaboration \cite{HERMES:2006jyl} as well as by the 
COMPASS collaboration \cite{Compass:2007qxf}.

The mass function used in Eq.(\ref{eq:dq}) is given as
\begin{align}\label{eq:mass}
g_{AV}(x) = \dfrac{1}{24b_x}\sqrt{x}(8-x-x^2)\tan^{-1}\left(\dfrac{2b_x}{\sqrt{x}}\right) - 
\dfrac{1}{24}x(2-(3+x)\log(x))
\end{align}
with $b_x = \sqrt{1-x/4}$.

\renewcommand{\theequation}{B.\arabic{equation}}
\setcounter{equation}{0}

\section*{Appendix B: Vertex Correction Evaluation}
\label{sec:App-B}

\begin{figure}[h!]
	\centering
	\includegraphics[scale=0.45]{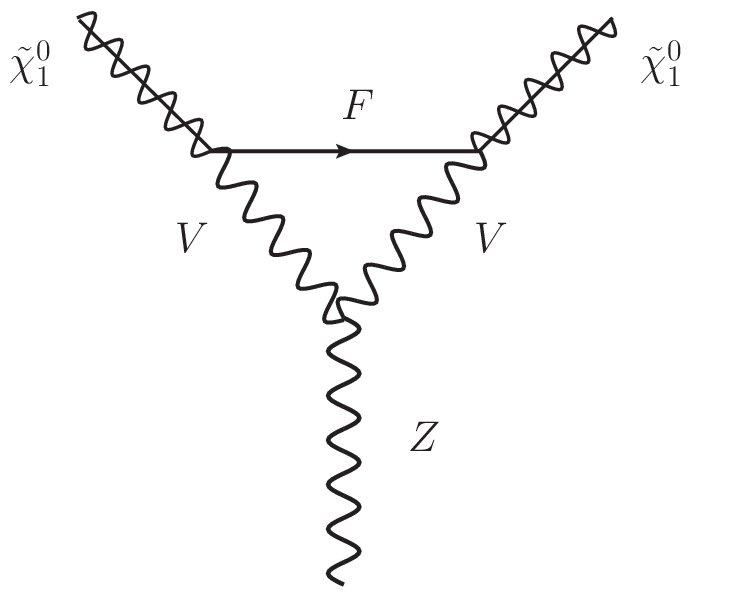}
	\includegraphics[scale=0.45]{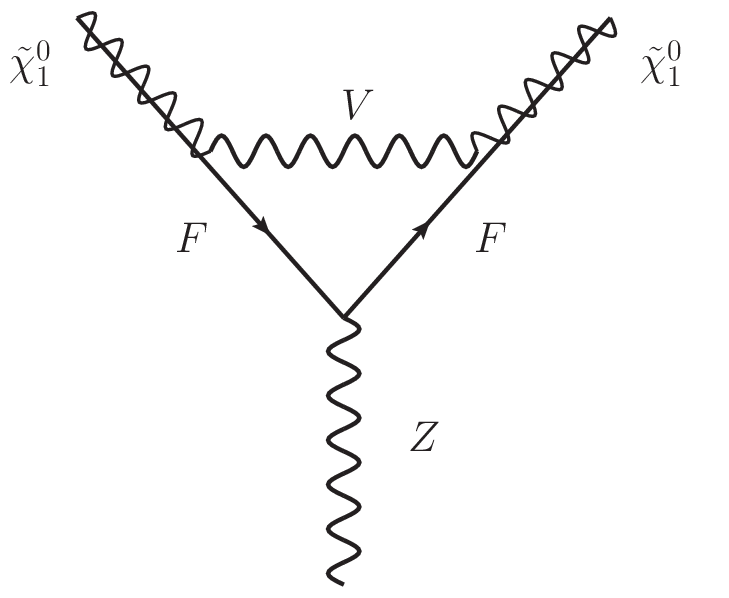}
	\includegraphics[scale=0.45]{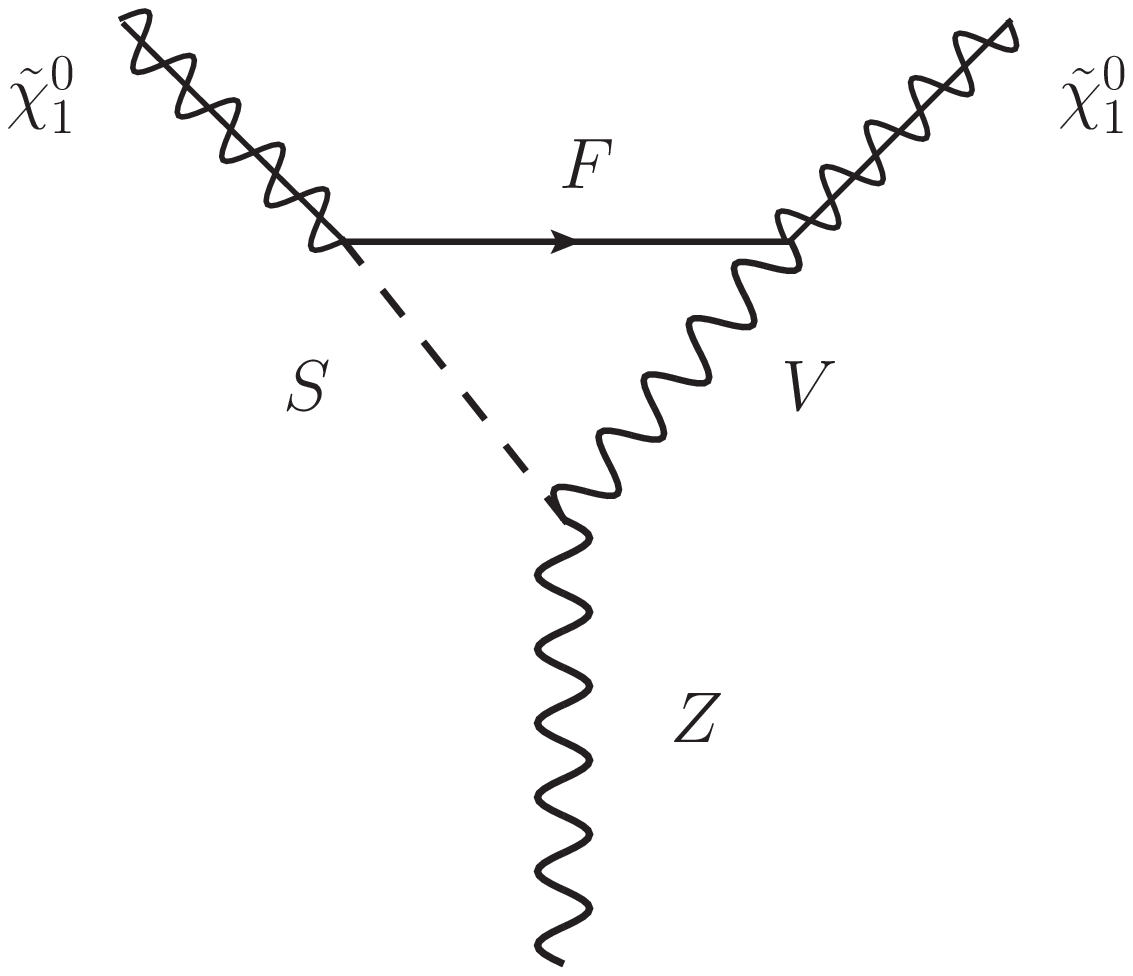} \\
	(a)\hspace{5.25cm}(b)\hspace{5.25cm}(c) \\
	\includegraphics[scale=0.45]{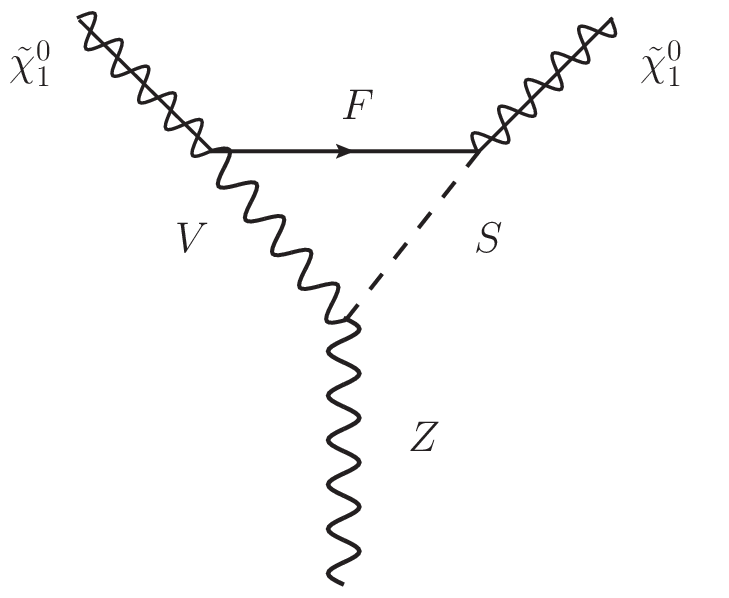}
	\includegraphics[scale=0.45]{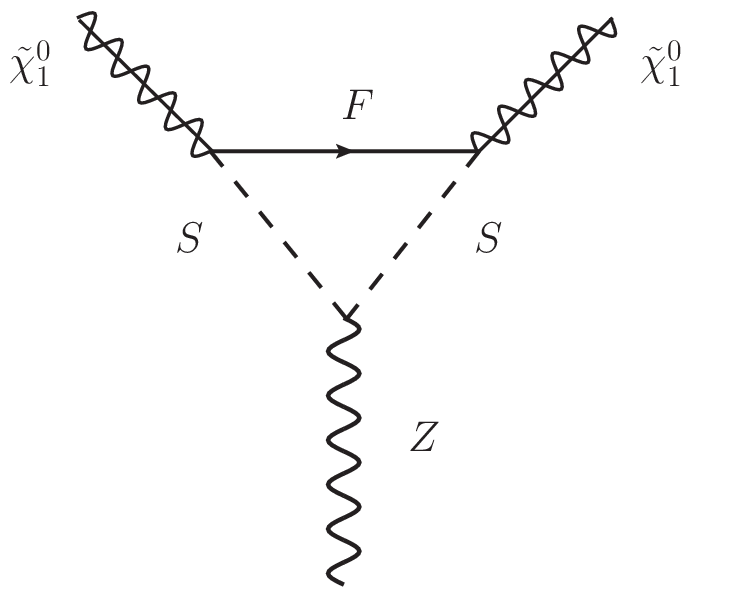}	
	\includegraphics[scale=0.45]{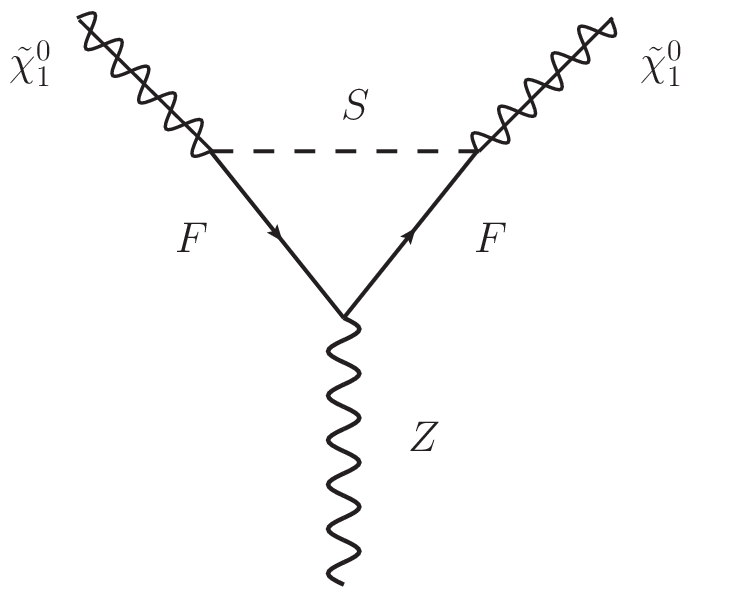}	\\
		(d)\hspace{5.25cm}(e)\hspace{5.25cm}(f) \\
	\caption{Topologies for the $\tilde{\chi}^0_1 \tilde{\chi}^0_1 Z$ vertex radiative correction 
	diagrams are shown in panels (a)-(f). Here, F represents fermions in the loop, V represents 
	vector bosons and S represents scalar particles.}
	\label{fig:Top1}
\end{figure}

\begin{figure}[h!]
	\centering
	\includegraphics[scale=0.6]{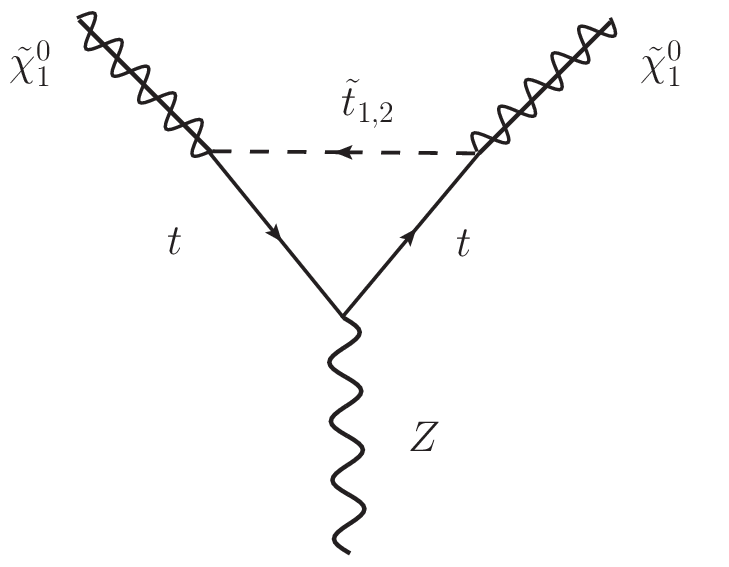}
	\includegraphics[scale=0.6]{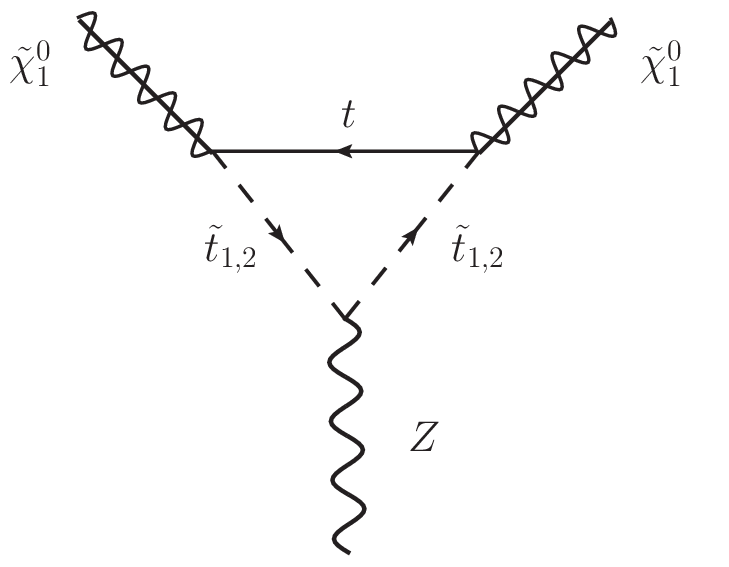}\\
	(a) \hspace{6.7cm} (b)
	\caption{Topology 1 {(panel(a)) and Topology 2 (panel (b)).}}
	\label{fig:vertexcorrtop}
\end{figure}

\begin{figure}[h!]
	\centering
	\includegraphics[scale=0.2]{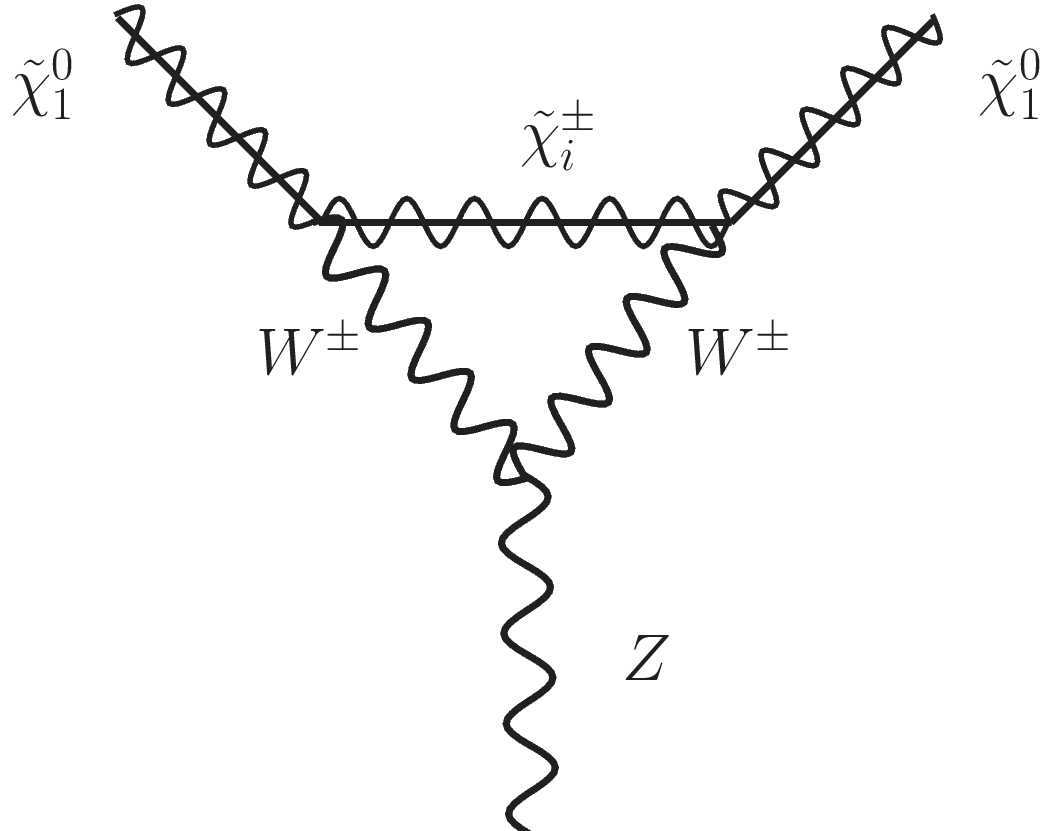}
	\includegraphics[scale=0.198]{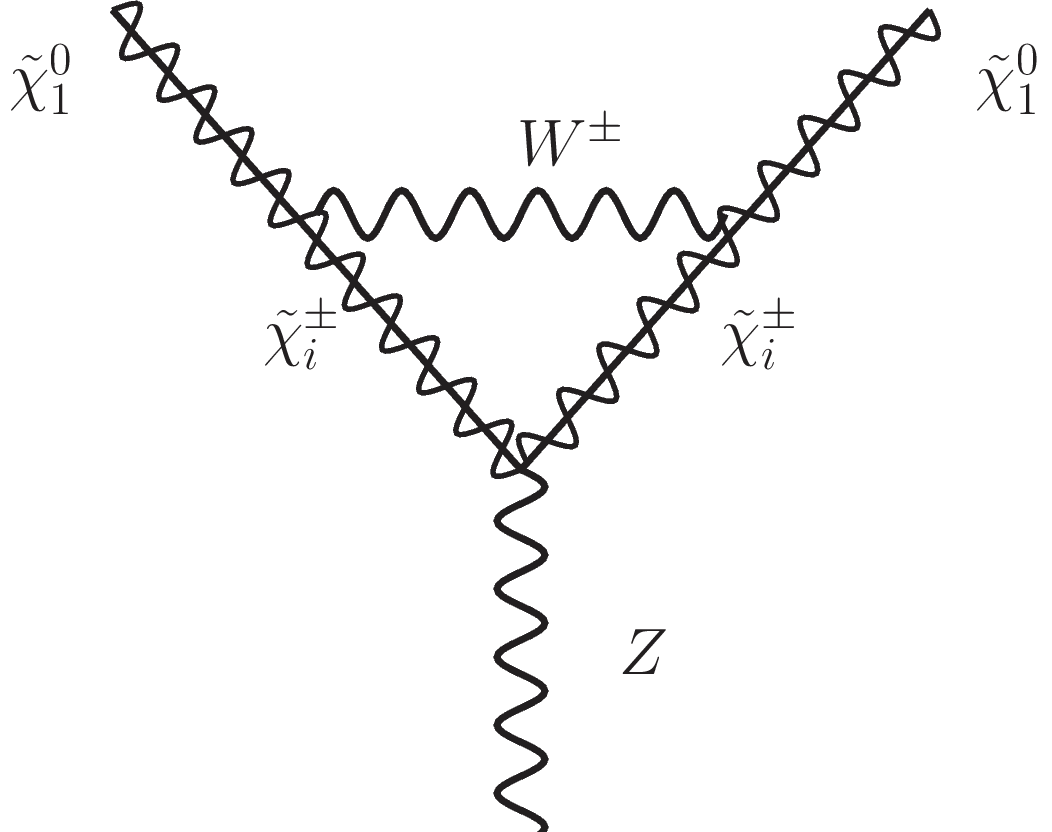}\\
	(a) \hspace{6.8cm} (b)	
	\caption{Topology 3 {(panel (a)) and Topology 4 (panel (b)).}}
	\label{fig:vertexcorrWW}
\end{figure}

There are 292 loop diagrams contributing to the vertex corrections, which can be categorized into 
six distinct topologies as depicted in Fig.\ref{fig:Top1}. For simplicity, we provide the explicit expressions for the four most dominant diagrams in our evaluation. The leading diagrams are 
illustrated in Fig.\ref{fig:vertexcorrtop} and Fig.\ref{fig:vertexcorrWW}, with their 
corresponding analytical expressions given in Eqs.(\ref{eq:top1}), (\ref{eq:top2}), 
(\ref{eq:WWCha}), and (\ref{eq:WCha2}). Contributions from the remaining diagrams are 
comparatively smaller and thus, their analytical expressions are omitted in this article. However, 
the expressions for these subdominant diagrams can be derived using the Feynman rules outlined in 
Ref.\cite{Drees:2004jm} or with the aid of \texttt{FeynArts} \cite{Hahn:2001rv}.

%\renewcommand{\theequation}{C.\arabic{equation}}
%\setcounter{equation}{0}
%\section*{Appendix C}

The mathematical equation for the Feynman diagram topology 1 in Fig.\ref{fig:vertexcorrtop}(a) is 
given as:
\begin{align}\label{eq:top1}
i\delta\Gamma^{(1)} =& \dfrac{i}{16 \pi^2} \biggl\{\gamma^{\mu} \mathbf{P_L}\left\{T_{1R} T_{3L}[m_t^2 T_{2R} C_0 -2 T_{2L} C_{00}-m_{\tilde{\chi}_1^0}^2 T_{2L}(C_{22}+2 C_{12}+ C_{11}+C_2+C_1)]\right\}\\ \nonumber
&+\gamma^{\mu} \mathbf{P_R}\left\{T_{1L} T_{3R}[m_t^2 T_{2L} C_0-2 T_{2R} C_{00}-m_{\tilde{\chi}_1^0}^2 T_{2R}(C_{22}+2 C_{12}+ C_{11}+C_2+C_1)]\right\}\biggr\} ,
\end{align}
where $C_i=C_i(m_{\tilde{\chi}^0_1}^2,q^2,m_{\tilde{\chi}^0_1}^2; m_{\tilde{t}^s}, m_t, m_t)$ and
$C_{ij}=C_{ij}(m_{\tilde{\chi}^0_1}^2,q^2,m_{\tilde{\chi}^0_1}^2; m_{\tilde{t}^s}, m_t, m_t)$
are the Passarino-Veltman functions \cite{PASSARINO1979151}. We have followed the 
convention of Refs.\cite{Patel:2015tea,vanOldenborgh:1990yc}. The explicit expressions for some of 
the relevant C functions are shown in the Appendix of Ref.\cite{Bisal:2023fgb}.
Further,
\begin{align*}
T_{1L}=& \dfrac{g_1[4 M_W s_{\beta} s_W \mathcal{U}_{\tilde{t}}(s,2)N_{11}-3 c_W m_t \mathcal{U}_{\tilde{t}}(s,1)N_{14}]}{3\sqrt{2} s_W s_{\beta} M_W}, \\
T_{1R}=& -\dfrac{g_1[M_W s_{\beta} \mathcal{U}_{\tilde{t}}(s,1)(s_W N_{11}+3 c_W N_{12})+3 c_W m_t 
\mathcal{U}_{\tilde{t}}(s,2)N_{14}]}{3\sqrt{2} s_W s_{\beta} M_W}, \\
T_{2L}=& \dfrac{2 g_1 s_W}{3}, \quad \quad T_{2R}= \dfrac{g_1 (1-4 c_W^2)}{6 s_W},\\
T_{3L}=& T_{1R}, \quad \quad\quad ~T_{3R}= T_{1L}.
\end{align*}
In the above equations (as well as the following ones), $i,j,s=1,2$, $s_W=\sin{\theta_W}$, $c_W=\cos{\theta_W}$, 
$s_{\beta}=\sin{\beta}$, and $\mathcal{U}_{\tilde{t}}(s,k)$ are the stop mixing matrix elements, 
where $k=1,2$.

Topology 2, Fig.\ref{fig:vertexcorrtop}(b):
\begin{align}\label{eq:top2}
i\delta\Gamma^{(2)} =& -\dfrac{i \xi}{16 \pi^2} \biggl\{\gamma^{\mu} \mathbf{P_L} [2 S_{1R} S_{2L} C_{00}]+\gamma^{\mu} \mathbf{P_R}[2 S_{1L} S_{2R} C_{00}]\biggr\},
\end{align}
where, $C_i=C_i(m_{\tilde{\chi}^0_1}^2,q^2,m_{\tilde{\chi}^0_1}^2; m_t, m_{\tilde{t}^s}, 
m_{\tilde{t}^s})$, and $C_{ij}=C_{ij}(m_{\tilde{\chi}^0_1}^2,q^2,m_{\tilde{\chi}^0_1}^2; m_t, m_{\tilde{t}^s}, m_{\tilde{t}^s})$. Further,
\begin{align*}
\xi=& \dfrac{g_1}{3 s_W} [(1-4 c_W^2)\mathcal{U}_{\tilde{t}^s}^2(s,1)+4 s_W^2 \mathcal{U}_{\tilde{t}^s}^2(s,2)],\\
S_{1L}=& \dfrac{g_1[M_W s_{\beta} \mathcal{U}_{\tilde{t}^s}(s,1)\{s_W N_{11}+3 c_W N_{12}\}+3 c_W m_t 
\mathcal{U}_{\tilde{t}^s}(s,2) N_{14}]}{3\sqrt{2} M_W s_{\beta} s_W},\\
S_{1R}=& \dfrac{g_1[-4 M_W s_{\beta} s_W \mathcal{U}_{\tilde{t}^s}(s,2)N_{11}+3 c_W m_t \mathcal{U}_{\tilde{t}^s}(s,1) N_{14}]}{3\sqrt{2} M_W s_{\beta} s_W},\\
S_{2L}=& -S_{1R}, \quad \quad S_{2R} = -S_{1L}.
\end{align*}

Topology 3, Fig.\ref{fig:vertexcorrWW}(a):
\begin{align}\label{eq:WWCha}
i\delta\Gamma^{(3)} =& \pm \dfrac{i g_2 c_W}{16 \pi^2} \biggl\{\gamma^{\mu} \mathbf{P_L}\left[2 Q_{2L}(-2 Q_{1L} C_{00}+m_{\tilde{\chi}_1^0}^2 Q_{1L}(C_2-C_1)+
m_{\tilde{\chi}^0_1} m_{\tilde{\chi}_i^{\pm}} Q_{1R}(2C_0+C_2-C_1))\right] \\ \nonumber
&+\gamma^{\mu} \mathbf{P_R}\left[2Q_{2R}(m_{\tilde{\chi}_1^0} m_{\tilde{\chi}_i^{\pm}} Q_{1L}(2 C_0 + C_2-C_1)-2 Q_{1R} C_{00}+m_{\tilde{\chi}_1^0}^2 Q_{1R}(C_2-C_1))\right] \biggr\},
\end{align}
where, $C_j=C_j(m_{\tilde{\chi}^0_1}^2,q^2,m_{\tilde{\chi}^0_1}^2; m_{\tilde{\chi}_i^{\pm}}, M_W, 
M_W)$, and $C_{jk}=C_{jk}(m_{\tilde{\chi}^0_1}^2,q^2,m_{\tilde{\chi}^0_1}^2; 
m_{\tilde{\chi}_i^{\pm}}, M_W, M_W)$. Further,
\begin{align*}
Q_{1L} =& -g_2 \left(\pm {V}_{i1} N_{12} - \dfrac{{V}_{i2} N_{14}}{\sqrt{2}}\right), \quad\quad 
Q_{1R} = -g_2 \left(\pm U_{i1} N_{12} + \dfrac{U_{i2} N_{13}}{\sqrt{2}} \right),\\
Q_{2L} =& Q_{1L}, \hspace{4.7cm}
Q_{2R}= Q_{1R}.
\end{align*}
From Eq.(\ref{eq:WWCha}) it is evident that the {loop diagrams involving 
($W^+,\tilde{\chi}_i^+$) and the ($W^-,\tilde{\chi}_i^-$) in Fig.\ref{fig:vertexcorrWW}(a),} 
contribute destructively and there is a partial cancellation between the positive and the 
negative {charge} diagrams. This leads to a small contribution 
from this diagram in the Higgsino case, {as this cancellation is not exact.} In the 
pure Higgsino limit {(and the pure Wino limit), this cancellation is exact and} the 
contribution from this diagram vanishes {for both the symmetric and anti-symmetric 
cases of the Higgsino-like (and Wino-like) benchmarks}.

Topology 4, Fig.\ref{fig:vertexcorrWW}(b):
\begin{align}\label{eq:WCha2}
i\delta\Gamma^{(4)} =& \pm \dfrac{i}{16 \pi^2} \biggl\{\gamma^{\mu} \mathbf{P_L} \Bigl[F_{2L}\{F_{1L} O_L(4C_{00}+m_{\tilde{\chi}_1^0}^2(2C_{22}+2(2C_{12}+C_{11})+6C_0+8C_2+8C_1)) \\ \nonumber
&-2F_{1L} O_R m_{\tilde{\chi}_i^{\pm}}^2 C_0\} \Bigr]+ \gamma^{\mu}\mathbf{P_R}\Bigl[F_{2R}\{-2F_{1R}O_L m_{\tilde{\chi}_i^{\pm}}^2 C_0+F_{1R}
O_R(4C_{00}+ \\ \nonumber
&m_{\tilde{\chi}_1^0}^2(2C_{22}+2(2C_{12}+C_{11})+6C_0+8C_2+8C_1))\}\Bigr]\biggr\} ,
\end{align}
where, $C_j=C_j(m_{\tilde{\chi}^0_1}^2,q^2,m_{\tilde{\chi}^0_1}^2; M_W,
m_{\tilde{\chi}_i^{\pm}},m_{\tilde{\chi}_i^{\pm}})$, and $C_{jk}=C_{jk}(m_{\tilde{\chi}^0_1}^2,q^2,m_{\tilde{\chi}^0_1}^2; M_W,
m_{\tilde{\chi}_i^{\pm}},m_{\tilde{\chi}_i^{\pm}})$. Further,
\begin{align*}
F_{1L} =& -g_2\left(\pm V_{i1}N_{12}-\dfrac{V_{i2}N_{14}}{\sqrt{2}}\right), \quad\quad \quad
F_{1R}= -g_2\left(\pm U_{i1}N_{12}+\dfrac{U_{i2}N_{13}}{\sqrt{2}} \right) ,\\
F_{2L} =& F_{1L}, \hspace{5.1cm} F_{2R} = F_{1R},\\
O_L =& \dfrac{g_2}{2 c_W} (-2 s_W^2 + 2|V_{11}|^2 + |V_{12}|^2), \quad\quad
O_R = \dfrac{g_2}{2 c_W} (-2 s_W^2 + 2|U_{11}|^2 + |U_{12}|^2).
\end{align*}
Similar to the previous case, in Eq.(\ref{eq:WCha2}) we see a partial cancellation between 
the diagrams {involving} ($W^+,\tilde{\chi}_i^+$) and the ($W^-,\tilde{\chi}_i^-$) 
states {in Fig.\ref{fig:vertexcorrWW}(b) resulting in a relatively diminished 
contribution especially in the Higgsino case. This cancellation is inexact, leading to a 
non-zero contribution in our case. In the pure Higgsino limit (and the pure Wino one), this is 
an exact cancellation and this loop diagram vanishes as well for both the symmetric and 
anti-symmetric cases of the Higgsino-like (and Wino-like) benchmarks.}

\bigskip
\bibliographystyle{JHEPCust.bst}
\bibliography{Higgsino}

\end{document}